\documentclass[pra,twocolumn,preprintnumbers,amsmath,amssymb]{revtex4}

\usepackage{graphicx}
\usepackage{dcolumn}
\usepackage{bm}
\usepackage{color}
\usepackage{epstopdf}
\usepackage{ulem}

\usepackage{ulem}
\def\eps{\varepsilon}

\newcommand{\br}{ {\bm r}}

\def\eps{\varepsilon}

\def\br{{\bm r}}

\begin{document}



\title{Bridging Rayleigh-Jeans and Bose-Einstein condensation of a guided fluid of light with positive and negative temperatures}
\author{L. Zanaglia$^1$, J. Garnier$^{2}$, S. Rica$^{3}$, R. Kaiser$^{1}$,  S. Wabnitz$^{4}$, C. Michel$^{1,5}$, V. Doya$^{1}$, A. Picozzi$^{6}$}
\affiliation{$^{1}$ Universit\'e C\^ote d'Azur, CNRS, Institut de Physique de Nice, Nice, France}
\affiliation{$^{2}$ CMAP, CNRS, Ecole Polytechnique, Institut Polytechnique de Paris, 91128 Palaiseau Cedex, France}
\affiliation{$^{3}$ 
Instituto de F\'isica, Pontificia Universidad Cat\'olica de Chile, Avenida Vicu\~na Mackenna 4860, Santiago, Chile}
\affiliation{$^{4}$ Department of Information Engineering, Electronics and Telecommunications, Universit\'a degli Studi di Roma Sapienza, Via Eudossiana 18, Rome 00184RM, Italy}
\affiliation{$^{5}$ Institut Universitaire de France (IUF), 1 rue Descartes, 75005 Paris, France}
\affiliation{$^{6}$ Laboratoire Interdisciplinaire Carnot de Bourgogne, CNRS, Universit\'e de Bourgogne, Dijon, France}


\begin{abstract}
We consider the free propagation geometry of a light beam (or fluid of light) in a multimode waveguide. As a result of the effective photon-photon interactions, the photon fluid thermalizes to an equilibrium state during its conservative propagation. In this configuration, Rayleigh-Jeans (RJ) thermalization and condensation of classical light waves have been recently observed experimentally in graded index multimode optical fibers characterized by a 2D parabolic trapping potential.
As well-known, the properties of RJ condensation differ substantially from those of Bose-Einstein (BE) condensation: The condensate fraction decreases quadratically with the temperature for BE condensation, while it decreases linearly for RJ condensation. Furthermore, for quantum particles the heat capacity tends to zero at small temperatures, and it takes a constant value in the classical particle limit at high temperatures. This is in contrast with classical RJ waves, where the specific 
heat takes a constant value at small temperatures, and tends to vanish above the condensation transition in the normal (uncondensed) state. 
Here, we reconcile the thermodynamic properties of BE and RJ condensation: By introducing a frequency cut-off inherent to light propagation in a waveguide, we derive generalized expressions of the thermodynamic properties that include the RJ and BE limits as particular cases. We extend the approach to encompass negative temperatures. In
contrast to positive temperatures, the specific heat does not display a singular behavior at negative temperatures, reflecting the non-critical nature of the transition to a macroscopic population of the highest energy level. Our work contributes to understanding the quantum-to-classical crossover in the equilibrium properties of light, within a versatile experimental platform based on nonlinear optical propagation in multimode waveguides.
\end{abstract}


\maketitle

\section{Introduction}


There is a current growing interest in studying both classical and quantum fluids of light \cite{carusotto13,bloch22,glorieux23,wrightNP22,ferraro23}.
We consider in this article the cavity-less configuration, where the fluid of light is a quasi-monochromatic beam that freely propagates in a conservative (Hamiltonian) nonlinear medium of the Kerr type.
The nonlinear photon-photon interactions can be viewed as the photonic counterpart to weakly interacting atomic Bose gases \cite{carusotto15,glorieux23}.
Indeed, fluids of light rely on the formal analogy between a laser field propagating through a nonlinear medium and the temporal evolution of a 2D quantum fluid.
In this way, 2D fluid of lights have been used to probe various phenomena, such as, e.g., the generation of superfluid Bogoliubov sound waves \cite{fontaine18,michel18,michel21,glorieux23b,braidotti22}, analogue gravity and cosmology \cite{glorieux22}, complex vortex dynamics 
\cite{fleischerBKT,congy24}, precondensation \cite{PRL18}, or the dynamical formation of prethermalized equilibrium states \cite{berges04,larre16,abuzarli22}.
It is in this configuration that Bose-Einstein (BE) condensation and thermalization of photons have been predicted theoretically in a conservative free propagation geometry  of a beam of light \cite{chiocchetta16}.
Note that other forms of condensation processes have been studied in optical cavity systems \cite{carusotto13,bloch22}, either in the genuine quantum BE regime \cite{weitz10,weitz10b,weitz16,nyman18,barland21,fischer19}, or in the classical regime 
\cite{conti08,PRA11b,berloff13,weill10,turitsyn12,turitsyn15,fratalocchi15,PRA15}, with specific  nonequilibrium features leading to different forms of universal properties \cite{zamora17}.


From a different perspective, several studies based on the wave turbulence theory \cite{zakharov92,newell01,nazarenko11,Newell_Rumpf,shrira_nazarenko13,during06} predict that nonlinear waves can exhibit a phenomenon of Rayleigh-Jeans (RJ) thermalization and condensation  \cite{nazarenko05,PRL05,nazarenko03,onorato06,berloff07,PD09,brachet11,
residori09,laurie12,PR14,aceves22,onorato23,nazarenko14,nazarenko23a,nazarenko23b}.
Although the physics of BE condensation and RJ condensation can be viewed differently, the underlying mathematical origin is similar, because of the common singular behavior (vanishing denominator) of the equilibrium Bose distribution for quantum particles, and the equilibrium RJ distribution for classical waves \cite{nazarenko11,PRL05}.

In the context of optics, RJ thermalization of light waves propagating in nonlinear media has been studied theoretically and experimentally for a long time \cite{PRL06,OE09,PRL10,NP12,EPL07,PRA13}.
However, it should be stressed that the thermalization to the RJ distribution is not properly defined, because it leads to diverging expressions of the optical power (`number of particles' in a corpuscular picture) and the kinetic energy of the beam.
This problem, known as the ultraviolet catastrophe of classical waves, can be regularized by  introducing a frequency cut-off. 
The introduction of a frequency cut-off is commonly used to develop a classical field description of quantum Bose gases for highly occupied modes \cite{davis01,blakie05,blakie08}, in line with the fact that the BE distribution recovers the classical RJ distribution for these modes.
On the other hand, an effective physical frequency cut-off was proposed in Ref.\cite{PRA11} by considering the propagation of optical waves in a waveguide configuration.
Considering the example of a multimode fiber (MMF) exhibiting a parabolic beam trapping potential (i.e., a graded-index MMF), it was shown theoretically that, owing to the Kerr nonlinearity, the optical field thermalizes through its propagation in the MMF toward the RJ mode power equilibrium distribution \cite{PRA11}.
The finite number of modes of the MMF then introduces a natural frequency cut-off, so that the process of RJ thermalization turns out to be properly defined, with converging expressions of both the energy and number of particles.
Recently, the observation of RJ thermalization has been reported experimentally in graded-index MMFs \cite{PRL20,mangini22,pourbeyram22,podivilov22,ferraro24}, as an explanation for the effect of spatial beam cleaning \cite{krupa16,krupa17,wright16,podivilov19}, along with its thermodynamic interpretation \cite{PRL19,PRA19,sidelnikov19,christodoulides19,parto19,makris20,
makris22,EPL21,kottos20,kottos23,
mangini23,kottos24,ferraro23_beam_beam,wabnitz23,OC23} -- see the recent review article \cite{ferraro23}.
In related more recent works, it has been pointed out that, for higher-order modes, the output mode power distribution that accompanies the beam self-cleaning experiments of \cite{mangini22,pourbeyram22} can be fitted by the BE law, see Refs.\cite{zitelli23}, as confirmed by recent experiments in long spans (up to 1km) of graded-index fibers \cite{zitelli23,zitelli24}.
As originally recognized in Ref.\cite{christodoulides19} for multimode optical systems, this observation is consistent with the fact that the RJ distribution can be viewed as a limiting case to the BE distribution for highly populated modes.
From a different perspective, a generalized optical thermodynamics approach has been discussed in \cite{steinmeyer23}, which ultimately suggests the BE distribution as the  natural convergence limit of light thermalization and condensation in MMFs.
This raises interesting questions about the relevance of the BE distribution to describe light thermalization in MMFs.

On the other hand, different numerical simulations have argued that a system of classical oscillators can equilibrate, in certain regimes, to a quantum BE distribution \cite{shepelyansky13}.
This puzzling issue has been clarified recently in Ref.\cite{shepelyansky23} by extensive numerical simulations based on the random matrix theory, which showed that the RJ distribution (instead of the BE distribution) plays the role of  universal attractor for weakly nonlinear systems.

It should be stressed that the thermodynamic equilibrium properties derived from the BE and RJ equilibria are of fundamentally different nature.
Let us consider the example of a 2D parabolic trapping potential that is relevant to the optical fiber experiments, where RJ thermalization has been recently reported \cite{PRL20,mangini22,pourbeyram22}.
In the BE case, the transition to condensation is characterized  by a condensate fraction that decreases quadratically with the temperature $\sim (T/T_c)^2$, where $T_c$ is the critical temperature of the condensation transition \cite{stringari16}.
In contrast, according to the RJ equilibrium, the condensate fraction decreases linearly with the temperature $\sim T/T_c$, as confirmed experimentally in Ref.\cite{PRL20}.
Even more surprising are the properties of the specific heat, which is known to be an important quantity for characterizing the phase transition to condensation \cite{stringari16}.
In the  BE case, the quantum property of frozen degrees of freedom imposes $C_V \to 0$ at small temperatures.
This is in contrast with the RJ distribution, where at small temperatures the heat capacity tends to a constant ($C_V \to M$, with $M$ the number of fiber modes), as a result of the theorem of energy equipartition inherent to classical statistical mechanics.
Conversely, at large temperatures, $C_V$ tends to a constant in the classical particle limit of the BE distribution \cite{stringari16}, whereas $C_V \to 0$ for guided RJ waves \cite{PRL20}.
It turns out that, so far, there is no evident connection that is established between RJ condensation observed in MMFs and the well-known equilibrium properties of BE condensation.

Our aim in this article is to clarify some of the puzzling issues discussed above.
Our main result is to bridge the thermodynamic equilibrium properties of RJ and BE condensation.
By introducing a frequency cut-off inherent to light propagation in a waveguide, we derive generalized theoretical expressions that unify the RJ and the BE equilibrium properties.
Our generalized expressions of the condensate fraction and specific heat include the RJ and BE regimes as particular cases, and thus explain the crossover between classical RJ condensation and quantum BE condensation.
In this way, we elucidate that in typical MMF experiments where the number of photons involved greatly exceeds the number of modes, light condensation is characterized by the classical RJ distribution.
Furthermore, the bounded energy spectrum inherent to light propagation in a waveguide configuration entails the existence of negative temperature equilibrium states \cite{onsager49,ramsey56,baldovin21}, a property  originally pointed out for classical multimode optical systems in Ref.\cite{christodoulides19}. 
Here, we extend the negative temperature RJ equilibrium states observed recently in MMFs \cite{PRL23} to the BE photon regime, and analyze the peculiar quantum features of the specific heat for negative temperatures.
From a broader perspective, this work contributes to the understanding of the quantum-to-classical transition in the framework of a flexible experimental platform based on the free propagation of beams of light in nonlinear media, such as optical fibers \cite{PRL20,mangini22,pourbeyram22,PRL22}, atomic vapors \cite{PRL18,PRA21,fontaine18,glorieux22,abuzarli22,glorieux23,congy24}, or photorefractive crystals \cite{fleischerBKT,michel18,michel21}.

\section{BE and RJ equilibriums}
\label{sec:BE_RJ_eq}

We recall that achieving complete spatial thermalization of a beam of light throughout its propagation in a nonlinear medium requires a guided configuration. 
Indeed, a non-confined speckled beam would experience a significant diffraction during its propagation, which would lead to a significant reduction of the power at the beam center. 
This inevitably affects the process of thermalization, whose equilibrium properties are only properly defined in the presence of a trapping potential.

Aside from these technical aspects, the presence of an inhomogeneous trapping potential is known to significantly affect the properties of the phase transition to condensation \cite{bagnato87}.  
In the following we consider the concrete example of a parabolic (graded-index) MMF, in which RJ thermalization has been observed and studied \cite{PRL20,mangini22,pourbeyram22,PRL23,ferraro24}. 
As will be discussed below in detail, we anticipate that the parabolic trapping potential plays a key role in 2D, since it enables a true phase transition to condensation in the thermodynamic limit, which is in contrast with a 2D uniform trapping potential (i.e., step-index optical waveguide) \cite{stringari16}.
Accordingly, the term ``condensation" refers here to a phase transition that occurs below a critical value of the temperature, and that is characterized by a macroscopic population of the fundamental mode.
More precisely, we consider a parabolic potential $V(\br)=q |\br|^2$ in 2D, that is truncated at $V_0=q R^2$, where $R$ is the fiber radius and $q$ (in m$^{-3}$) a constant determined by the fiber characteristics.
The guided mode eigenvalues are well approximated by the ideal harmonic potential $\beta_p=\beta_0(p_x+p_y+1)$ (in m$^{-1}$), 
where $p$ labels the two integers $(p_x,p_y)$  that specify a mode.
The regular spacing among groups of degenerate modes is $\beta_0=\sqrt{2q/k_0}$, with $k_0=2 \pi n_0/\lambda$, $\lambda$ being the laser wavelength.
The number $G$ of groups of degenerate modes (number of `energy levels') in the truncated parabolic potential, and the corresponding number of modes $M$, are given by
\begin{equation}
G=V_0/\beta_0, \qquad M=G(G+1)/2 \simeq G^2/2.
\label{eq:G_M}
\end{equation}
We assume throughout the paper that $G \gg 1$ (highly multimode waveguide).

The number of photons $n_p$ at thermal equilibrium in the mode $p=(p_x,p_y)$ is given by the BE statistics \cite{chiocchetta16}:
\begin{align}
n_p^{BE} = \frac{1}{\exp\Big( \frac{\eps_p - \mu}{k_B T}\Big) - 1},
\end{align}
where $T$ (in K) and $\mu$ (in J) are the physical temperature and the chemical potential, respectively, $k_B$ being the Boltzmann's constant.
The energy per particle in the mode $p$ is $\eps_p = \hbar c_0 \beta_p$ (in J), with $c_0=c/n_0$ the group-velocity of light in the fiber ($n_0$ being the core refractive index), and $\hbar$ the Planck's constant.
This expression of $\eps_p$ can be obtained by writing the temporal evolution equation for the quantum electric field operator in the Heisenberg picture, see Ref.\cite{carusotto15} for details.
In order to make a link with the RJ distribution, it proves convenient to define an `optical temperature' ${\tilde T}=k_B T / (\hbar c_0)$ (in m$^{-1}$), as well as an `optical chemical potential' ${\tilde \mu}=\mu / (\hbar c_0)$ (in m$^{-1}$), so that 
\begin{align}
n_p^{BE} = \frac{1}{\exp\Big( \frac{\beta_p - {\tilde \mu}}{{\tilde T}}\Big) - 1}.  
\label{eq:BE}
\end{align}
The recent experiments in MMFs \cite{PRL20,mangini22,pourbeyram22} have reported the observation of light thermalization to the RJ equilibrium distribution:
\begin{align}
n_p^{RJ}=\frac{{\tilde T}}{\beta_p - {\tilde \mu}}. 
\label{eq:RJ}
\end{align}
Note that, the BE distribution recovers the RJ distribution in the limit where the modes are highly populated \cite{zakharov92,christodoulides19,davis01,blakie05,blakie08}.
The total number of photons is $N = \sum_p n_p$ and the corresponding total kinetic energy $E = \sum_p \varepsilon_p n_p$.
We define in the following the `optical energy' ${\tilde E}=E/(\hbar c_0)=\sum_p \beta_p n_p$ (in m$^{-1}$).

Before proceeding, it is important to recall that thermalization to equilibrium does not necessarily imply a phenomenon of condensation.
A well-known example is provided by the black body radiation, in which the number of photons $N$ is not conserved, so that the chemical potential is zero in the Bose distribution, $\mu=0$ \cite{huang}.
Consequently, the Bose distribution no longer describes a phase transition toward a macroscopic population of the fundamental mode -- the photon gas exhibits thermalization, but not condensation, see Refs.\cite{weitz10,weitz10b}.
In a similar way, there exist a large number of classical wave systems (capillary waves, acoustic waves, Rossby planetary waves \cite{nazarenko11}, or vibrating elastic plates \cite{during06}) that do not conserve the `number of particles' $N$, so that $\mu=0$, and condensation does not take place.
In other words, the equilibrium distribution does not exhibit a singular behavior, and the fundamental mode does not become macroscopically populated, i.e.,  $n_0^{eq}$ is of the same order as any other $n_{p \neq 0}^{eq}$, whatever the energy (or the temperature).

It is also noteworthy that, in the strong nonlinear regime, the condensed component forms a distinct phase characterized by unique properties compared to the thermal (uncondensed) component.
For instance, the presence of a strong condensate modifies the linear dispersion relation of the thermal component, so that the condensate amplitude $n_0^{eq}$ is not simply given by the RJ or BE distributions (\ref{eq:BE}-\ref{eq:RJ}) taken at $p=0$. One needs to resort to a Bogoliubov-like transformation~\cite{bogoliubov47} in order to properly describe the condensate fraction at equilibrium, see Ref.\cite{stringari16} and Ref.\cite{PRL05} for the quantum and classical approaches, respectively. 
The modified Bogoliubov dispersion relation has been recently measured experimentally in an optical system~\cite{fontaine18}.
It is at the origin of well distinguished physical properties that have been  investigated experimentally in fluids of light, such as superfluidity \cite{frisch1992transition,carusotto2014superfluid,michel18,braidotti22}, or  the turbulence flow of superfluid light past an obstacle  \cite{michel21,glorieux23b}, in relation with quantum vortex dynamics and the Berezinskii-Kosterlitz-Thouless transition \cite{fleischerBKT}.

In contrast to the strong nonlinear regime of condensation, so far, the experiments on RJ thermalization in MMFs have been carried out in the weakly nonlinear regime \cite{PRL20,mangini22,pourbeyram22,podivilov22,ferraro24}.
As is well-known from the weak turbulence theory \cite{zakharov92,newell01,nazarenko11,PLA10}, in the weak nonlinear regime the formation of large-scale coherent structures (vortices, solitons, shock or rogue waves,...) is essentially ruled out.
More precisely, it was shown in Ref.\cite{EPL21} that the Bogoliubov nonlinear renormalization of the dispersion relation can be neglected in the experiments of RJ thermalization \cite{PRL20,mangini22,pourbeyram22,podivilov22,ferraro24}, and that the condensate is stable with respect to the focusing nonlinearity in MMFs.


The weakly nonlinear hypothesis has permitted the development of a kinetic wave turbulence description of the nonequilibrium process of RJ thermalization observed in MMFs \cite{PRA11,PRL19,PRA19,PRL22}. 
Indeed, in the weakly nonlinear regime, the nonlinear (interaction) contribution to the energy can be neglected, so that the linear energy ${\tilde E}$ refers to the conserved linear Hamiltonian \cite{PRA11}.
The equilibrium properties can then be investigated in the framework of the ideal gas approximation, by neglecting  nonlinear interactions.
Accordingly, $N$ and ${\tilde E}$ are conserved quantities during light propagation in the nonlinear medium, so that the optical temperature and chemical potential $({\tilde T}, {\tilde \mu})$ at equilibrium (i.e., at the fiber output) are determined uniquely by the conserved number of particles and optical energy ($N, {\tilde E}$), an important property that has been rigorously proved only recently in Ref.\cite{parto19} on the basis of Bolzano's theorem.
In other terms, there is no thermostat in the optical experiments \cite{PRL20,mangini22,pourbeyram22}: By keeping constant the number of particles $N$, the energy ${\tilde E}$ plays the role of the control parameter in the transition to condensation for this microcanonical statistical ensemble.
Here, we first follow the traditional treatment of BE condensation, where the temperature is the control parameter of the phase transition to condensation. 
Next, in section~\ref{sec:n_0vsE}, we convert the expressions of the condensate fraction as a function of the energy. 

The starting point of our analysis is the BE distribution with a frequency cut-off, which, as discussed above, has its origin in the finite number of modes inherent to light propagation in the waveguide configuration.   
Next we study different limits in terms of the mode occupations: 
In the limit of weak mode occupations, we recover the equilibrium properties known for the BE case, whereas in the limit of strong mode occupations, we recover the equilibrium properties known for the RJ case.

\section{Condensate fraction vs temperature}

\subsection{Density of states and thermodynamic limit}
\label{sec:dos_TL}

In order to derive analytically tractable expressions and thereby gain physical insight into the relationship between BE and RJ condensation, 
in the following we introduce the so-called semi-classical approximation \cite{stringari16}, where the discrete sums involved in the number of particles ($N=\sum_p n_p$) and the energy (${\tilde E}=\sum_p \beta_p n_p$) can be replaced by continuous integrals.
As is well-known, the passage from  discrete to continuous sums is justified when 
specific conditions are met.
The number of photons must be large $N \gg 1$, as well as the number of groups of modes, $G \gg 1$. In addition, the equilibrium distribution must populate a large number of mode groups. This is usually verified when the optical energy is large, typically ${\tilde E} \gg {\tilde E}_0$, where ${\tilde E}_0=N\beta_0$ is the minimum energy when all photons populate the fundamental mode.
This latter condition can also be expressed in terms of an optical temperature, which should be much larger than the spacing levels of the parabolic potential, ${\tilde T} \gg \beta_0$ (or equivalently $k_B T \gg \hbar \varepsilon_0$) \cite{stringari16}.
The density of states reads \cite{note_dos}:
\begin{align}
\rho(\beta) = \frac{\beta}{\beta_0^2}   \quad {\rm for} \quad \beta \le V_0,
\label{eq:dos}
\end{align}
and $\rho(\beta) = 0$ for $\beta > V_0$. 
Aside from its truncation at $V_0$ that is inherent to the waveguide configuration, this expression of the density of states coincides with that of Ref.\cite{stringari16}.
Note in particular that the number of modes is approximated by 
\begin{align*}
M=\sum_p^{G} 1 \simeq \int_0^{V_0} \rho(\beta) d\beta=G^2/2,
\end{align*}
where $\sum_p^{G}=\sum_{0\le p_x+p_y < G}$ denotes the sum truncated by the finite number of modes of the waveguide.
Also, the number of particles and the optical energy read
\begin{align}
N=\int_0^{V_0} \rho(\beta) n(\beta) d\beta, \qquad 
{\tilde E}=\int_0^{V_0} \rho(\beta) \beta n(\beta) d\beta,
\label{eq:N_E_continTL}
\end{align}
where $n(\beta)$ stands for either the BE or the RJ distributions
\begin{align}
n^{BE}(\beta)=\frac{1}{e^{\frac{\beta-{\hat \mu}}{{\tilde T}}}-1}, 
\qquad 
n^{RJ}(\beta)=\frac{{\tilde T}}{\beta-{\hat \mu}},
\label{eq:BE_RJ_n_beta}
\end{align}
with the shifted chemical potential ${\hat \mu}=\mu-\beta_0$.
In these expressions the continuous variable $\beta \in [0, V_0]$.
We can, however, observe that a singularity of the BE and RJ distributions at $\beta=0$ can occur  when ${\hat \mu} = 0$. This means that the semi-classical approximation needs to be updated in such a situation \cite{stringari16}. The discrete sums involved in the number of particles $N$ and the energy $\tilde{E}$ should then  be decomposed in discrete components supported on the fundamental mode and continuous components that represent the contributions of the other modes. This will give an updated version of Eq.~(\ref{eq:N_E_continTL}): we are going to study this situation in the next section. 

The passage from discrete sums to continuous integrals is also important to study the thermodynamic limit.
In the presence of a parabolic confining potential, the density of photons is $N\beta_0^2$ (in m$^{-2}$) \cite{stringari16}, so that the thermodynamic limit is defined as 
$N\to \infty$, $\beta_0 \to 0$, while keeping constant 
\begin{equation}
N \beta_0^2={\rm const},   \qquad  V_0={\rm const}.
\label{eq:TL}
\end{equation}
It has been shown in Ref.\cite{PRA11} that RJ condensation occurs in the thermodynamic limit in the presence of a 2D parabolic potential.
In addition, the experimental results of the transition to condensation reported in Ref.\cite{PRL20} were found to be close to the thermodynamic limit, despite the presence of finite size effects inherent to the real experiment.
In the following section, we will justify the approach by studying the convergence to the (continuous) thermodynamic limit.
We will consider the general case, as well as the RJ and BE limits.

\begin{figure}
\includegraphics[width=.8\columnwidth]{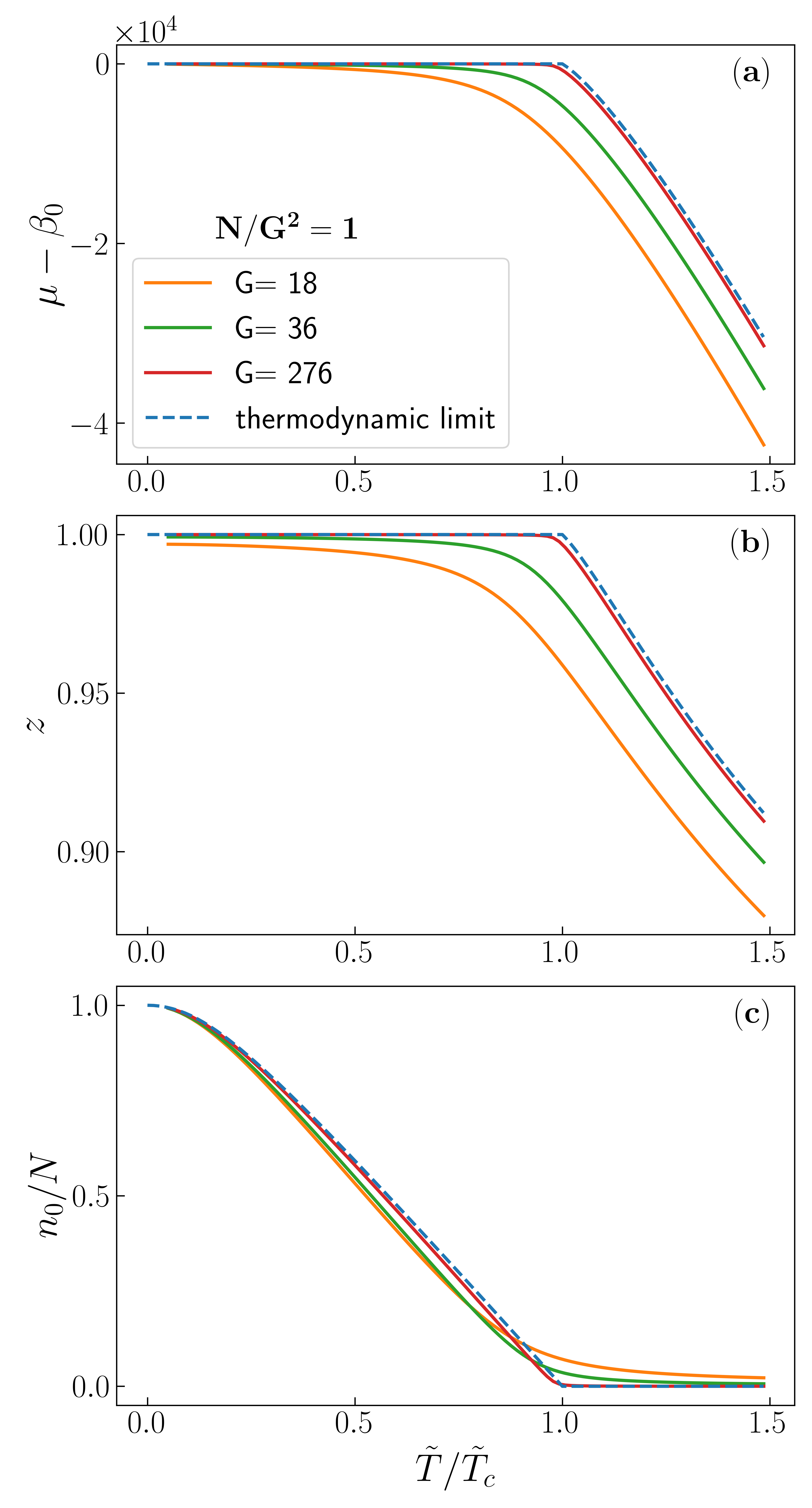}
\caption{
\baselineskip 10pt
{\bf Convergence to the  thermodynamic limit for $N/G^2=1$.} 
(a) Chemical potential vs temperature, ${\tilde \mu}({\tilde T})$; (b) fugacity vs temperature, $z=e^{({\tilde \mu}-\beta_0)/{\tilde T}}$; (c) condensate fraction vs temperature, $n_0({\tilde T})/N$.   
The solid lines refer to the computation of the discrete sums beyond the thermodynamic limit from Eqs.(\ref{eq:N_mu_T})-(\ref{eq:N_frac_bey_TL}).
By increasing $N$ and $G$ [while keeping $N/G^2=$const and $V_0=$const, see Eq.(\ref{eq:NsG2vsNbetsV0})], the system size increases and approaches the thermodynamic limit (dashed blue lines, from Eq.(\ref{eq:N_int_x_basic}) and Eq.(\ref{eq:N_frac})), where the curves ${\tilde \mu}({\tilde T})$ in (a), $z({\tilde T})$ in (b), and $n_0({\tilde T})/N$ in (c), display a singular cusped behavior at ${\tilde T} = {\tilde T}_c$ [solution of Eq.(\ref{eq:N_int_x})]. 
The intermediate regime is considered with $N/G^2=1$ (see Figs.~\ref{fig:0_RJ}-\ref{fig:0_BE} for $N/G^2 \gg 1$ and $N/G^2 \ll 1$, respectively). 
}
\label{fig:0} 
\end{figure}

\begin{figure}
\includegraphics[width=.8\columnwidth]{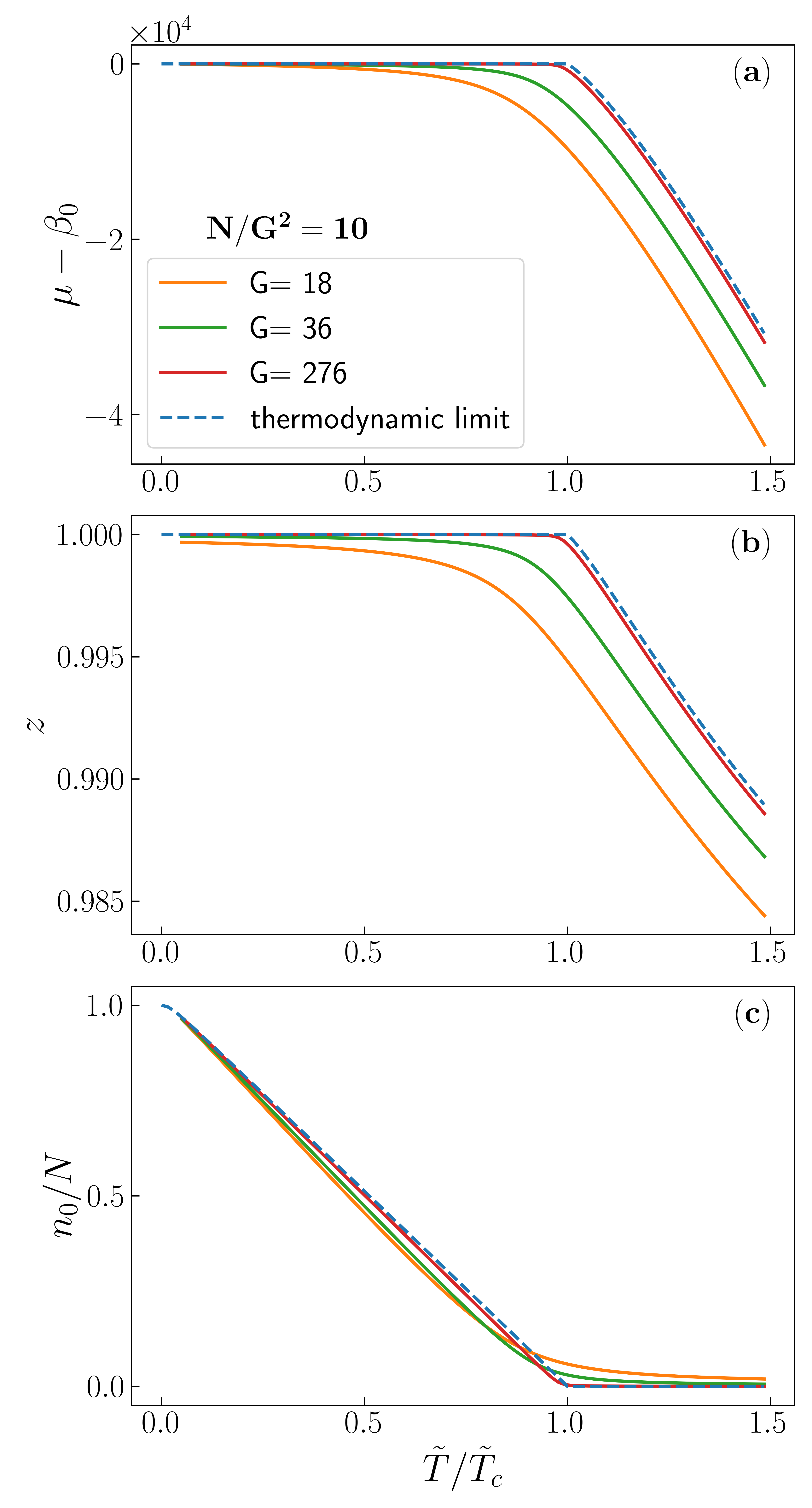}
\caption{
\baselineskip 10pt
{\bf Convergence to the  thermodynamic limit for $N/G^2 \gg 1$.} 
(a) ${\tilde \mu}({\tilde T})$; (b) $z({\tilde T})$; (c) $n_0({\tilde T})/N$.   
The solid lines refer to the computation of the discrete sums beyond the thermodynamic limit from Eqs.(\ref{eq:N_mu_T})-(\ref{eq:N_frac_bey_TL}).
By increasing $N$ and $G$ [while keeping $N/G^2=$const and $V_0=$const, see Eq.(\ref{eq:NsG2vsNbetsV0})], the system size increases and approaches the thermodynamic limit (dashed blue lines, from Eq.(\ref{eq:N_int_x_basic}) and Eq.(\ref{eq:N_frac})), where the curves ${\tilde \mu}({\tilde T})$ in (a), $z({\tilde T})$ in (b), and $n_0({\tilde T})/N$ in (c), display a singular cusped behavior at ${\tilde T} = {\tilde T}_c$ [solution of Eq.(\ref{eq:N_int_x})]. 
The RJ regime is considered with $N/G^2=10$, see Fig.~\ref{fig:RJ_vs_BE}. 
}
\label{fig:0_RJ} 
\end{figure}

\begin{figure}
\includegraphics[width=.8\columnwidth]{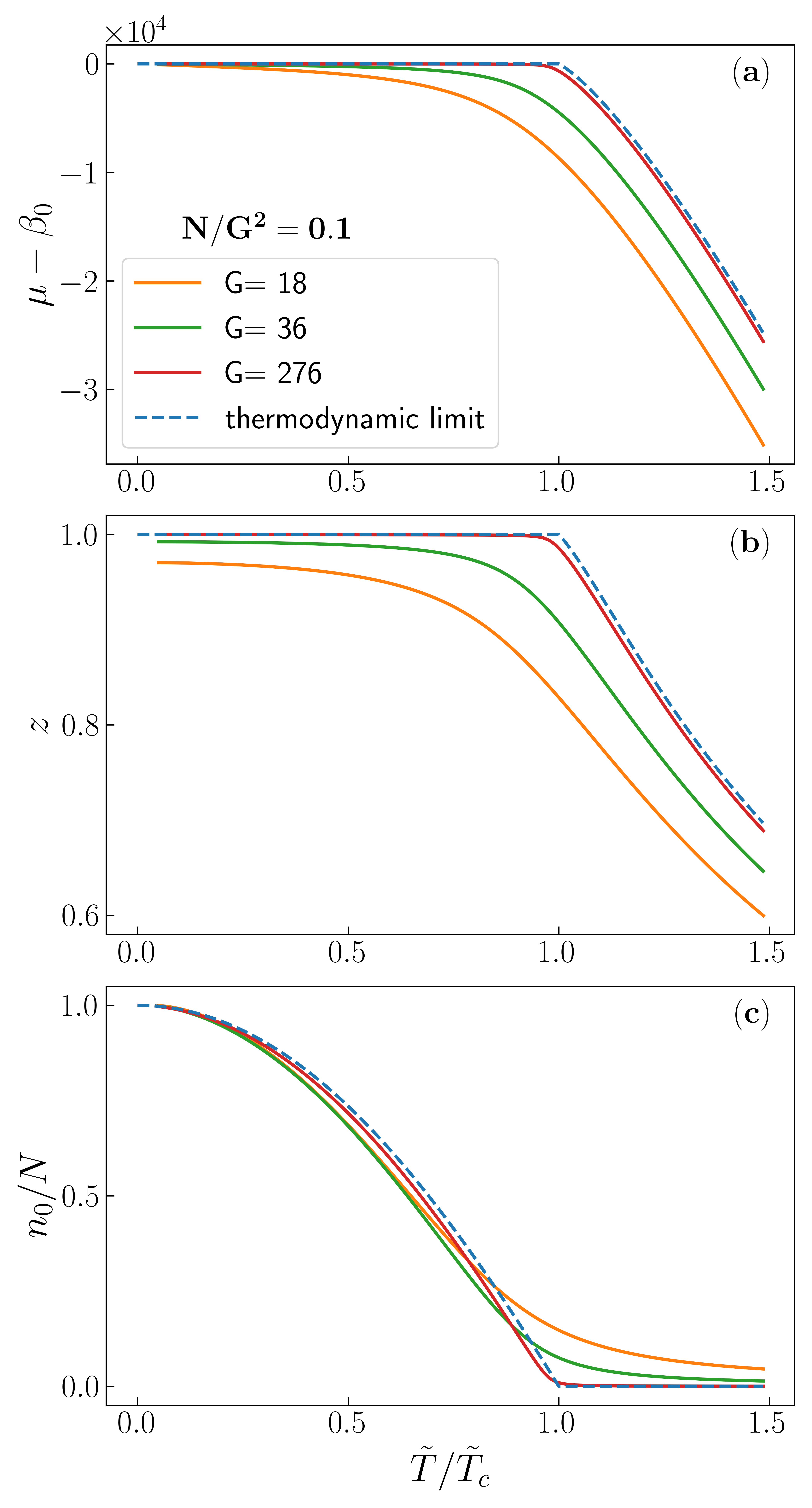}
\caption{
\baselineskip 10pt
{\bf Convergence to the  thermodynamic limit for $N/G^2 \ll 1$.} 
(a) ${\tilde \mu}({\tilde T})$; (b) $z({\tilde T})$; (c) $n_0({\tilde T})/N$.   
The solid lines refer to the computation of the discrete sums beyond the thermodynamic limit from Eqs.(\ref{eq:N_mu_T})-(\ref{eq:N_frac_bey_TL}).
By increasing $N$ and $G$ [while keeping $N/G^2=$const and $V_0=$const, see Eq.(\ref{eq:NsG2vsNbetsV0})], the system size increases and approaches the thermodynamic limit (dashed blue lines, from Eq.(\ref{eq:N_int_x_basic}) and Eq.(\ref{eq:N_frac})), where the curves ${\tilde \mu}({\tilde T})$ in (a), $z({\tilde T})$ in (b), and $n_0({\tilde T})/N$ in (c), display a singular cusped behavior at ${\tilde T} = {\tilde T}_c$ [solution of Eq.(\ref{eq:N_int_x})]. 
The BE regime is considered with $N/G^2=0.1$, see Fig.~\ref{fig:RJ_vs_BE}. 
}
\label{fig:0_BE} 
\end{figure}


\subsection{Condensate fraction: Convergence to the thermodynamic limit}
\label{sec:converg_TL}

\subsubsection{Beyond the thermodynamic limit}

From the formal point of view, quantum BE condensation \cite{stringari16}, and classical RJ condensation \cite{PRA11,PRL20}, originate in the singularity of the equilibrium distributions at ${\tilde \mu}=\beta_0$.
The physical behavior goes as follows: 
By keeping constant the number of photons $N$, and by decreasing the temperature ${\tilde T}$, the chemical potential ${\tilde \mu}$ increases and tends toward the fundamental mode eigenvalue from below, ${\tilde \mu} \to  \beta_0^-$.
This general behavior is illustrated in Fig.~\ref{fig:0}, in which we report the chemical potential vs the temperature obtained from the relation
\begin{align}
N = \sum_{p}^G \frac{1}{e^{\frac{\beta_p-{\tilde \mu}}{{\tilde T}}}-1}.
\label{eq:N_mu_T}
\end{align}
Each curve refers to fixed values of $(N,G,V_0)$, and then for a given temperature ${\tilde T}$, Eq.(\ref{eq:N_mu_T}) is solved to get the corresponding ${\tilde \mu}$, which gives ${\tilde \mu}({\tilde T})$.
In Fig.~\ref{fig:0}, the potential depth has been fixed to $V_0=181 \, 346$m$^{-1}$ for all curves. This particular value has been chosen because it corresponds to a commercially available graded-index MMF with a rather large number of mode groups $G \simeq 276$ ($\beta_0=657$m$^{-1}$).
Then the different curves reported in Fig.~\ref{fig:0} with an increasing value of $G$, correspond to an increase of the system size (i.e., an increase of the fiber radius), while keeping $V_0=$const.

The important point to note in Fig.~\ref{fig:0}(a), is that, as the system size increases and approaches the thermodynamic limit (\ref{eq:TL}), the chemical potential tends to  reach the fundamental mode eigenvalue at a specific critical temperature ${\tilde T}_c$ that will be defined below.
For completeness, we have also reported in Fig.~\ref{fig:0}(b) the corresponding behavior of the fugacity, defined by $z=e^{({\tilde \mu}-\beta_0)/{\tilde T}}$, which will be shown to play an important role in the subsequent analysis.

Let us now analyze the behavior of the condensate fraction.
At variance with Eq.(\ref{eq:BE_RJ_n_beta}), the condensate amplitude does not diverge beyond the thermodynamic limit, because the chemical potential is smaller than the fundamental mode eigenvalue, ${\tilde \mu} < \beta_0$, so that $n_0 = 1/[ e^{ (\beta_0-{\tilde \mu})/{\tilde T} }-1  ]$.
Equivalently, following the usual treatment of BE condensation \cite{stringari16}, the condensate fraction is computed by isolating $n_0$, that is 
$N=n_0 + \sum_{p \neq 0}^G n_p^{BE}$, which immediately gives 
\begin{align}
\frac{n_0}{N} = 1 - \frac{1}{N} \sum_{p \neq 0}^G \frac{1}{e^{\frac{\beta_p-{\tilde \mu}}{{\tilde T}}}-1}.
\label{eq:N_frac_bey_TL}
\end{align}
The relation ${\tilde \mu}({\tilde T})$ obtained above from Eq.(\ref{eq:N_mu_T}) can be substituted in Eq.(\ref{eq:N_frac_bey_TL}) to get $n_0/N$ vs ${\tilde T}$.
The corresponding curves of the condensate fraction vs temperature reported in Fig.~\ref{fig:0}(c) evidence that the transition to condensation becomes critical as the system approaches the  thermodynamic limit, as revealed by a singular behavior (discontinuous derivative) of $n_0({\tilde T})/N$ at ${\tilde T}= {\tilde T}_c$ in the thermodynamic limit (dashed blue line). 
Conversely, the singularity at the phase transition is smoothed out by finite size effects beyond the thermodynamic limit, see Fig.~\ref{fig:0}(c).

To complete our study, we also report in Figs.~\ref{fig:0_RJ}-\ref{fig:0_BE} the  convergence to the thermodynamic limit of the chemical potential, the fugacity, and the condensate fraction, for the regimes $N/G^2 \gg 1$ and $N/G^2 \ll 1$, which will be shown to correspond to the RJ and BE regimes,  respectively, see Fig.~\ref{fig:RJ_vs_BE}.

\subsubsection{Thermodynamic limit}

Let us now focus the analysis of the transition to condensation in the  thermodynamic limit.
As discussed above in section~\ref{sec:dos_TL}, in this limit the discrete sum in Eq.(\ref{eq:N_mu_T}) can be converted into a continuous integral, so that the number of photons (\ref{eq:N_E_continTL}) reads
\begin{align}
N = G^2 \int_0^{1} \frac{x}{z^{-1} \exp\big( \frac{V_0}{{\tilde T} }x\big) - 1} dx,
\label{eq:N_int_x_basic}
\end{align}
where we have introduced the change of variable $x=\beta/V_0$, and we recall that $z=e^{({\tilde \mu}-\beta_0)/{\tilde T}}$. 
For a set of values $(N,G,V_0)$, Eq.(\ref{eq:N_int_x_basic}) provides the temperature dependence of the chemical potential ${\tilde \mu}({\tilde T})$,  and of the fugacity $z({\tilde T})$, in the  thermodynamic limit, see Fig.~\ref{fig:0}(a)-(b) (dashed blue line).
The key observation is that, in this limit, the chemical potential exactly reaches the fundamental mode eigenvalue  ${\tilde \mu}= \beta_0$ (i.e., $z=1$) at a non vanishing critical temperature ${\tilde T}_c >0$, which is therefore defined by
\begin{align}
&\frac{N}{G^2} = \int_0^{1} \frac{x}{\exp\big( \frac{V_0}{{\tilde T}_c }x\big) - 1} dx
\nonumber\\
&= \Big( \frac{{\tilde T}_c}{V_0}\Big)^2 \Big[ \frac{V_0}{{\tilde T}_c} 
\log\Big( 1 - e^{-\frac{V_0}{{\tilde T}_c}}\Big) - 
g_2\Big( e^{-\frac{V_0}{{\tilde T}_c}}\Big) +\frac{\pi^2}{6}   \Big],
\label{eq:N_int_x}
\end{align}
where $g_p(z)=\frac{1}{\Gamma(p)}\int_0^\infty dx \frac{x^{p-1}}{z^{-1} e^x-1}=\sum_{l=1}^\infty \frac{z^l}{l^p}$ is the Bose function, with the Gamma function $\Gamma(p)=(p-1)!$ \cite{stringari16}. 

\begin{figure}
\includegraphics[width=.8\columnwidth]{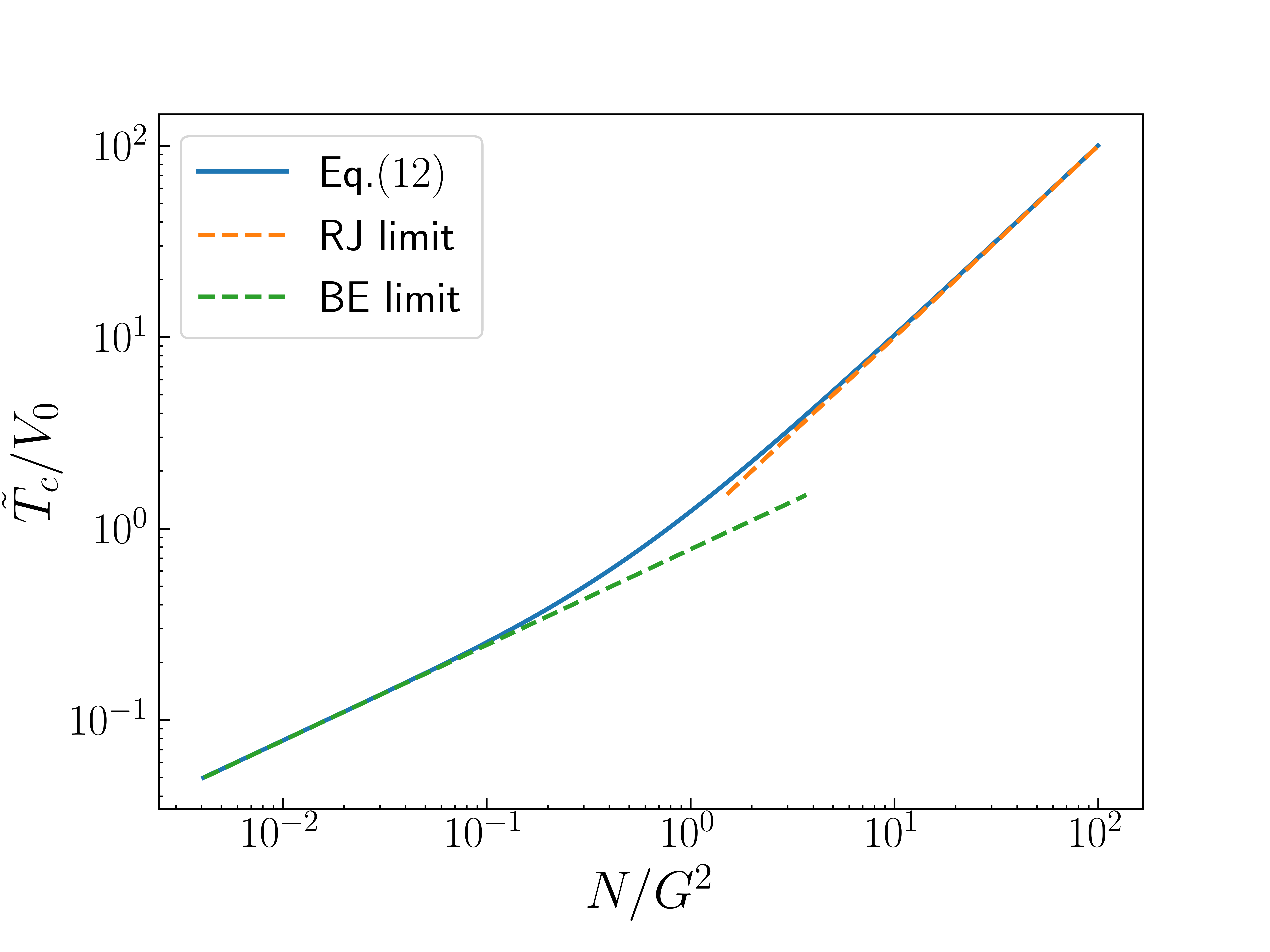}
\caption{
\baselineskip 10pt
{\bf Critical temperature.} 
$\tilde T_c/V_0 $ vs $N/G^2$: 
The blue line corresponds to the exact general expression given in Eq.(\ref{eq:N_int_x}). The green-dashed line reports Eq.(\ref{eq:Tc_BE}): It 
corresponds to the BE limit that is valid for $N/G^2 \ll 1$.
The orange-dashed line reports Eq.(\ref{eq:T_c_rj}): It corresponds to the RJ limit that is valid for $N/G^2 \gg 1$.
Note the square-root scaling, $\tilde T_c/V_0 \sim \sqrt{N/G^2}$, for the BE case; and the linear scaling, $\tilde T_c/V_0 \sim N/G^2$, for the RJ case.
}
\label{fig:RJ_vs_BE} 
\end{figure}

As discussed above through Eq.(\ref{eq:BE_RJ_n_beta}), in the continuous limit, the BE and RJ distributions exhibit a vanishing denominator for $\beta=0$ at ${\tilde \mu}= \beta_0$ (or ${\hat \mu}=0$). 
The singularity is treated by splitting the condensate and the thermal contribution \cite{stringari16}: $N=n_0+\int_0^{V_0} d\beta \rho(\beta)/[\exp(\beta/{\tilde T})-1]$ for ${\tilde T} \le {\tilde T}_c$.
The condensate fraction can then be written in the form:
\begin{align}
\frac{n_0}{N} = 1 - \frac{G^2}{N}\int_0^{1}  \frac{x}{\exp\big( \frac{V_0}{{\tilde T} }x\big) - 1} dx.
\label{eq:N_frac}
\end{align}
Note that, at ${\tilde T}={\tilde T}_c$ the condensate vanishes $n_0=0$, and (\ref{eq:N_frac}) recovers Eq.(\ref{eq:N_int_x}).
Actually, Eq.(\ref{eq:N_int_x}) determines the value of the dimensionless parameter ${\tilde T}_c/V_0$ as a function of $N/G^2$, see Fig.~\ref{fig:RJ_vs_BE} (blue line).
In practice, for a real experiment with finite size effects, the values of $N$, $G$ and $V_0$ are given. 
Note however that the quantity
\begin{equation}
N/G^2=N\beta_0^2/V_0^2
\label{eq:NsG2vsNbetsV0}
\end{equation} 
remains constant by increasing the system size up to the thermodynamic limit, see Eq.(\ref{eq:TL}).
The convergence to the thermodynamic limit of the condensate fraction is illustrated in Fig.~\ref{fig:0}(c) for the intermediate regime with $N/G^2=1$, as well as in panels (c) of Fig.~\ref{fig:0_RJ}-\ref{fig:0_BE} for the RJ regime ($N/G^2 \gg 1$), and BE regime ($N/G^2 \ll 1$), respectively.
Note in particular that in all regimes, the condensate fraction exhibits a singular behavior at the critical temperature ${\tilde T}_c$, which is smoothed out by finite size effects.

We remark that the convergence to the thermodynamic limit by increasing the system size is faster in the RJ regime than the BE regime.
Indeed, we will show below that in the BE regime ($N/G^2 \ll 1$), we have ${\tilde T}_c/V_0 \sim \sqrt{N/G^2}$ [see Eq.(\ref{eq:Tc_BE})].
Accordingly, in the BE regime the discrete sum (\ref{eq:N_mu_T}) can be approximated by an integral, with an error that scales as $\beta_0/{\tilde T}_c \sim 1/\sqrt{N}$. 
Conversely, in the RJ regime ($N/G^2 \gg 1$), we have ${\tilde T}_c/V_0 \sim N/G^2$ [see Eq.(\ref{eq:T_c_rj})], so that the discrete sum (\ref{eq:N_mu_T}) can be approximated by an integral, with an error that scales as $\beta_0/{\tilde T}_c \sim G/N$.
Consequently, the error resulting from approximating the discrete sum (\ref{eq:N_mu_T}) by an integral is much smaller in the RJ regime than in the BE regime,  since $G/N =\sqrt{G^2/N} / \sqrt{N} \ll 1/\sqrt{N}$.

In the following, we consider in our plots the commercially available graded-index MMF specified above through the Fig.~\ref{fig:0}, with the following parameters $G=276$, $V_0=181 \, 346$m$^{-1}$ and $\beta_0=657$m$^{-1}$ (numerical aperture NA=0.29).


\subsubsection{Uniform trapping potential (step-index MMF)}

Let us  comment the case of `step-index' MMFs, which are characterized by a homogeneous circular trapping potential, i.e., $V(\br)=0$ for $|\br|\le R$, and $V(\br)=V_0$ for $|\br| > R$, $R$ being the fiber core radius.
For a uniform trapping potential, the density of states in 2D is known to be a constant that does not depend on $\beta$: $\rho=k_0 A/(2\pi)$ for $\beta \leq V_0$ \cite{arendt}, where $A=\pi R^2$ is the area of the fiber core, and we recall that $k_0=2 \pi n_0/\lambda$.
Because the density of states is constant, the integrals in Eqs.(\ref{eq:N_int_x}-\ref{eq:N_frac}) involve $\int_0^1 1/[\exp(V_0 x/\tilde{T})-1]dx$, which no longer converges (the integrand is equivalent to $1/x$ at $x=0$).
Accordingly, the transition to condensation for a homogeneous trapping potential does not take place in the thermodynamic limit for a positive critical temperature, i.e., it occurs at zero temperature, ${\tilde T}_c=0$.

\begin{figure}
\includegraphics[width=.8\columnwidth]{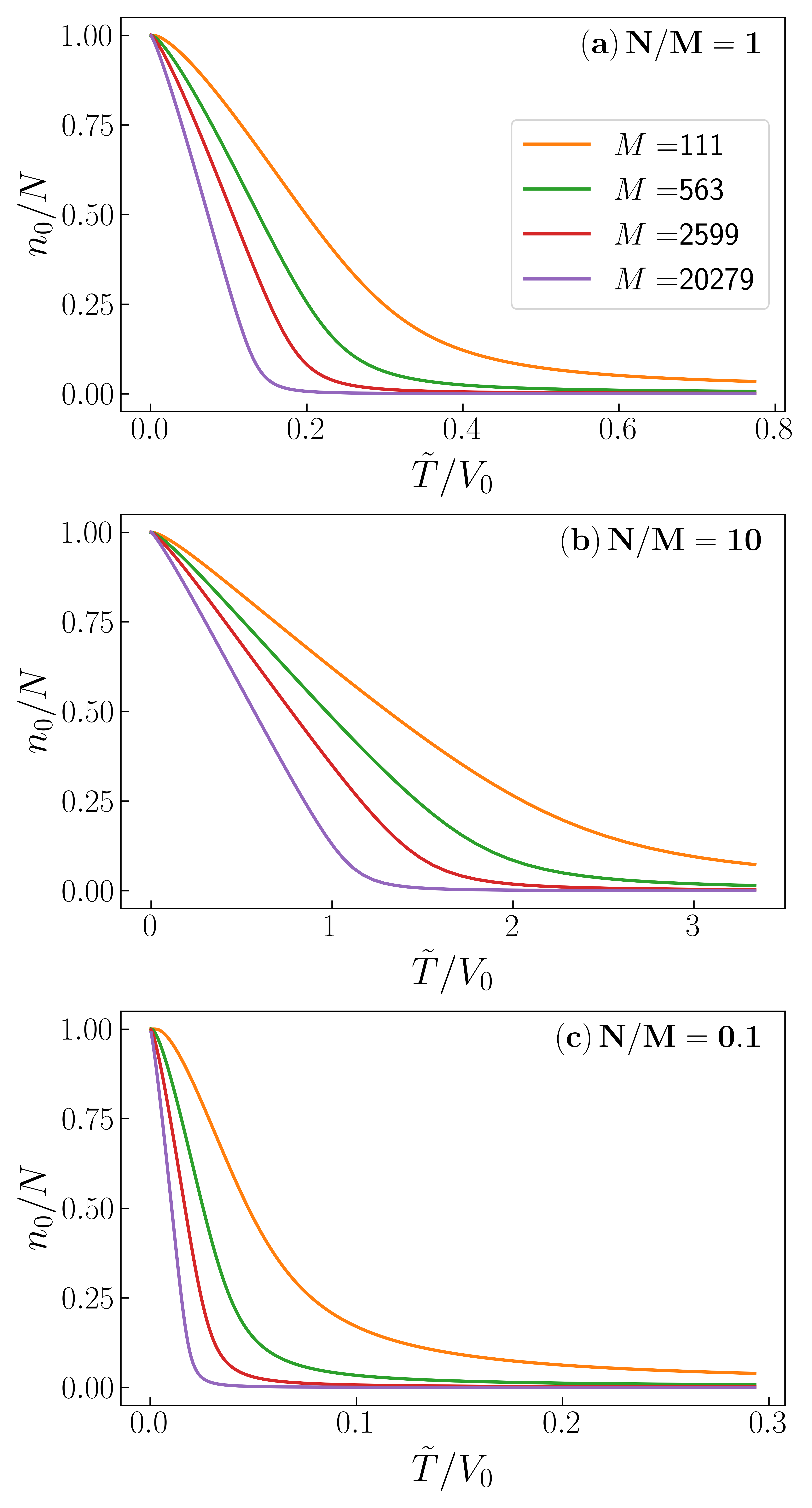}
\caption{
\baselineskip 10pt
{\bf Thermalization without condensation for a uniform trapping potential (step-index MMF).} 
Condensate fraction $n_0/N$ vs ${\tilde T}/V_0$ computed from Eq.(\ref{eq:N_frac_bey_TL}) beyond the thermodynamic limit, in the intermediate regime $N/M=1$ (a), the RJ regime $N/M=10$ (b), the BE regime $N/M=0.1$ (c). 
At variance with Figs.~\ref{fig:0}-\ref{fig:0_BE}, by increasing the system size (i.e., by increasing $N$ and $M$ while keeping $N/M=$const and $V_0=$const), the curves do not converge to a single cusped curve, because in the thermodynamic limit the critical temperature for condensation vanishes, ${\tilde T}_c=0$. 
Note that, for a given temperature ${\tilde T}>0$, the condensate amplitude decreases to zero in the  thermodynamic limit, $n_0({\tilde T}) \to 0$.
The photon gas exhibits thermalization to equilibrium, but there is no phase transition to condensation in the thermodynamic limit.
}
\label{fig:homog} 
\end{figure}

Let us analyze this aspect in more detail. For a 2D homogeneous system, the thermodynamic limit refers to the limits, $N \to \infty$ and $A \to \infty$, while keeping constant the particle density $N/A=$const and $V_0=$const. 
Note that the fiber core area is proportional to the number of modes, since $M=\int_0^{V_0} \rho d\beta= k_0 A V_0/(2\pi)$, so that the thermodynamic limit can be equivalently taken by letting $N \to \infty$  and $M \to \infty$, while keeping $N/M=$const and $V_0=$const. 
Accordingly, we report in Fig.~\ref{fig:homog} the condensate fraction vs temperature for a homogeneous circular waveguide in the intermediate regime ($N/M=1$), the RJ regime ($N/M=10$), and the BE regime ($N/M=0.1$).
The condensate fraction $n_0/N$ is computed from Eq.(\ref{eq:N_frac_bey_TL}), which involves a discrete sum over the modes, and it is thus valid beyond the thermodynamic limit \cite{note_beta_ls}.
We remark in Fig.~\ref{fig:homog} that the behaviors of the condensation curves are in  contrast with those discussed above through Figs.~\ref{fig:0}-\ref{fig:0_BE} for the parabolic trapping potential (graded-index MMF): As the area $A$ of the system increases, the curves in Fig.~\ref{fig:homog} do not tend to converge to a single cusped curve at some  positive critical temperature ${\tilde T}_c >0$. 
The key observation in Fig.~\ref{fig:homog} is that, for a given temperature ${\tilde T}$, the condensate fraction $n_0/N$ decreases to zero as the size of the system increases.
In other words, the observation of a macroscopic population of the fundamental mode is a manifestation of finite-size effects, which tends to disappear as the system size increases.
In this way, the photon gas exhibits a process of thermalization to equilibrium, but not condensation.
It is indeed well-known that there is no phase transition to condensation in a 2D homogeneous trapping potential, neither for RJ waves \cite{PRA11}, nor BE particles \cite{stringari16}.
For this reason, we do not consider in the following the case of step-index MMFs, and focus our analysis on graded-index MMFs.

\subsection{Condensate fraction: BE limit}

The expression of the condensate fraction given in Eq.(\ref{eq:N_frac}) is exact and valid in general.
Let us now consider the BE and RJ condensation limits.
As we shall see, the BE limit is obtained when the total number of
photons is much smaller than the number of fiber modes.
Indeed from (\ref{eq:N_int_x}) we have $N/G^2 = (\tilde{T}_c/V_0)^2 \int_0^{V_0/\tilde{T}_c} x/[\exp(x)-1] dx$. If $N/G^2 \ll 1$, then $\tilde{T}_c/V_0 \ll 1$ and therefore $N/G^2 \simeq (\tilde{T}_c/V_0)^2 \int_0^\infty x/[\exp(x)-1] dx$ (because the integral is convergent):
\begin{align}
N = G^2 \int_0^{\infty} \frac{x}{\exp\big( \frac{V_0}{{\tilde T}_c }x\big) - 1} dx = G^2 \zeta(2) \Big( \frac{{\tilde T}_c}{V_0} \Big)^2,
\label{eq:NG2_Tc_BE}
\end{align}
where $\zeta(p)=g_p(1)= \sum_{l=1}^\infty 1/l^p$ is the Riemann $\zeta-$function.
Then for $N \ll M$, Eq.(\ref{eq:N_frac}) recovers the expression of the BE condensate fraction without a frequency cut-off, that is 
\begin{align}
\frac{n_0^{BE}}{N} = 1 - \Big(\frac{{\tilde T}}{{\tilde T}_c}\Big)^2,
\label{eq:n0_BE}
\end{align}
with the critical temperature ${\tilde T}_c= \beta_0 \sqrt{N/\zeta(2)}$ \cite{stringari16}.
Note that this expression can also be written 
\begin{align}
{\tilde T}_c/V_0 = \sqrt{N/G^2}/\zeta(2).
\label{eq:Tc_BE}
\end{align}
This expression is reported in Fig.~\ref{fig:RJ_vs_BE}, see the dashed green line.

\begin{figure}
\includegraphics[width=.8\columnwidth]{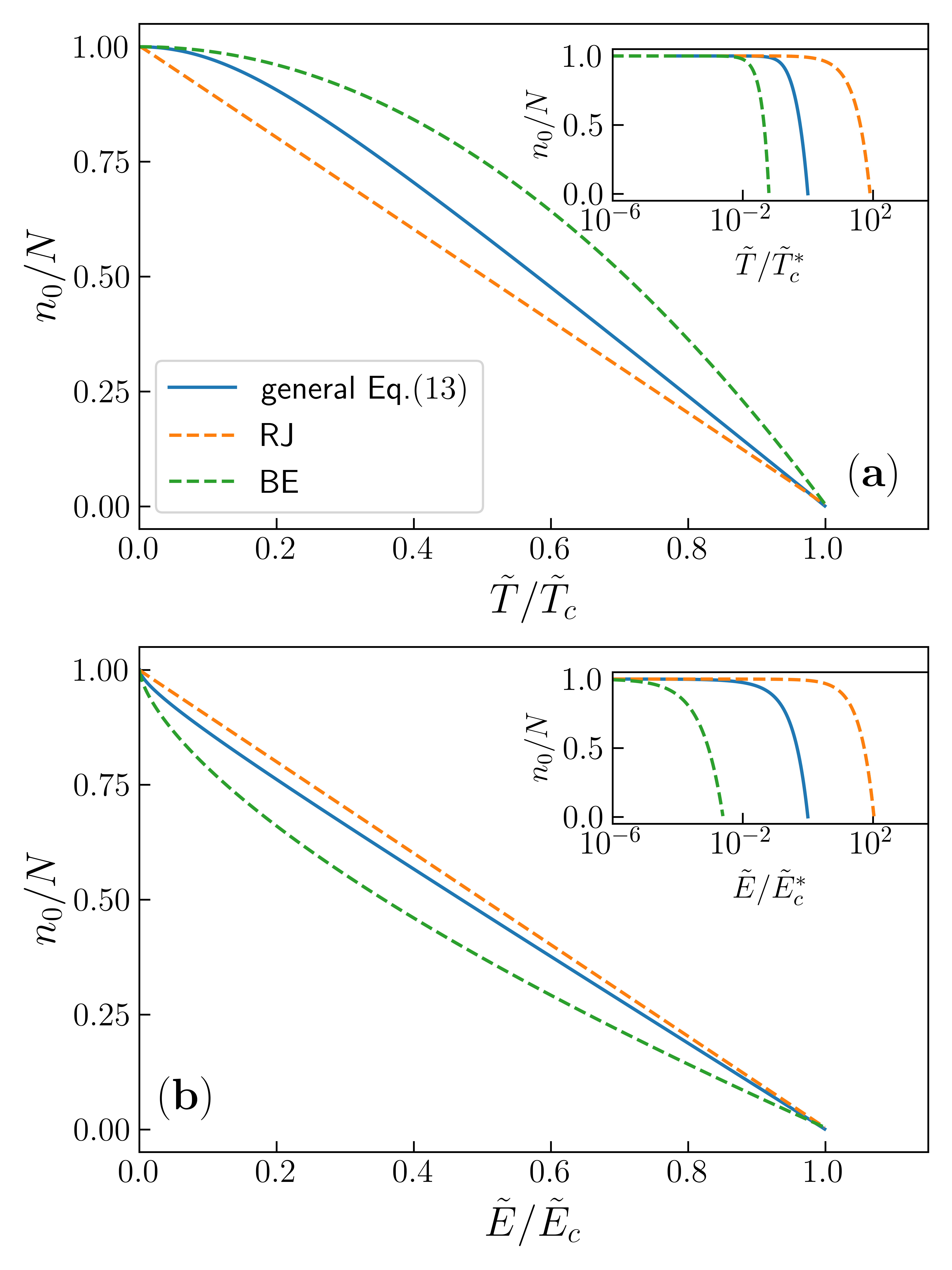}
\caption{
\baselineskip 10pt
{\bf Condensate fraction.} 
(a) $n_0/N$ vs ${\tilde T}/{\tilde T}_c$ obtained from Eq.(\ref{eq:n0_BE}) in the BE limit (dashed green), and from Eq.(\ref{eq:n0_RJ}) in the RJ limit (dashed orange).
In the intermediate case (solid blue), it is obtained from the exact general expression Eq.(\ref{eq:N_frac}) with $N/G^2=1$: Note the hybrid behavior, which is almost quadratic at small temperatures ${\tilde T}/{\tilde T}_c \ll 1$, and almost linear nearby the critical temperature ${\tilde T}/{\tilde T}_c \lesssim  1$.
(b) Corresponding condensate fraction as a function of the energy: $n_0/N$ vs ${\tilde E}/{\tilde E}_c$ is obtained from Eq.(\ref{eq:cond_BE}) in the BE limit, from Eq.(\ref{eq:cond_RJ}) in the RJ limit, and in the intermediate case with $N/G^2=1$.
Note that, for clarity, the temperature in (a) has been scaled with respect to the different critical temperatures ${\tilde T}_c$, respectively in the BE regime [see Eq.(\ref{eq:Tc_BE})], in the RJ regime [see Eq.(\ref{eq:T_c_rj})], and in the general case [from Eq.(\ref{eq:N_int_x})] -- the same has been done in (b) with the critical energies ${\tilde E}_c$ in the BE, RJ, and in the general case.
For completeness, we report in the insets the same curves by scaling the temperature (energy) with respect to the same critical temperature ${\tilde T}_{c}^*$ (energy ${\tilde E}_{c}^*$) corresponding to $N/G^2=1$ (note the log-scale in the horizontal axis).
}
\label{fig:n_0vsT_E} 
\end{figure}


\subsection{Condensate fraction: RJ limit}

The RJ limit is obtained in the opposite case, whenever the total number of photons is much larger than the number of guided modes.
Indeed, if $N \gg G^2$, then we have ${\tilde T}_c \gg V_0$ by (\ref{eq:N_int_x})
and we can expand 
$\exp\big( \frac{V_0}{{\tilde T}_c }x\big) \simeq 1+ V_0x/{\tilde T}_c$, 
so that Eq.(\ref{eq:N_int_x}) takes the form 
\begin{align}
{\tilde T}_c/V_0=N/G^2.
\label{eq:T_c_rj}
\end{align}
The RJ regime then corresponds to the case $N/M \simeq {\tilde T}_c/V_0 \gg 1$, where $M \simeq G^2/2$ is the number of fiber modes.
Eq.(\ref{eq:T_c_rj}) is reported in Fig.~\ref{fig:RJ_vs_BE}, see the orange dashed line.
The critical temperature for RJ condensation is then ${\tilde T}_c = N V_0/G^2$, and the condensate fraction takes the form:
\begin{align}
\frac{n_0^{RJ}}{N} = 1 - \frac{{\tilde T}}{{\tilde T}_c}.
\label{eq:n0_RJ}
\end{align}
Referring back to the thermodynamic limit discussed here above through Eq.(\ref{eq:TL}), it becomes apparent that the transition to condensation takes place for a non-vanishing positive critical temperature in the thermodynamic limit, i.e., ${\tilde T}_c=N \beta_0^2/V_0 >0$.

\subsection{Discussion}

We report in Fig.~\ref{fig:n_0vsT_E}(a) the condensate fraction vs temperature in the BE limit $N/G^2 \ll 1$, which is characterized  by a quadratic behavior, and in the RJ limit $N/G^2 \gg 1$, where it exhibits a linear behavior.
The condensate fraction in the intermediate case for $N/G^2=1$ is also reported from the exact general expression given by Eq.(\ref{eq:N_frac}), which exhibits a hybrid quadratic-linear behavior, at small and  at high temperatures, respectively.

To summarize, the optical beam thermalizes in the classical RJ condensation regime for ${\tilde T}_c/V_0 \gg 1$, and quantum effects start to play a role when ${\tilde T}_c$ becomes of the order of $V_0$.
More precisely, for ${\tilde T}_c=V_0$ we have 
\begin{align*}
N = G^2 \int_0^{1} \frac{x}{\exp( x) - 1} dx \simeq 0.777 \times G^2 \simeq M, 
\end{align*}
and quantum condensation effects arise when the photon number $N$ decreases, and approaches the number of modes $M$.
This is illustrated in Fig.~\ref{fig:RJ_vs_BE} that reports  $\tilde T_c/V_0$ vs $N/G^2$ for the exact general case [Eq.(\ref{eq:N_int_x})], while the BE limit [Eq.(\ref{eq:Tc_BE})] and the RJ limit [Eq.(\ref{eq:T_c_rj})], provide the corresponding approximations for $N/G^2 \ll 1$ and  $N/G^2 \gg 1$, respectively.

\begin{figure}
\includegraphics[width=.8\columnwidth]{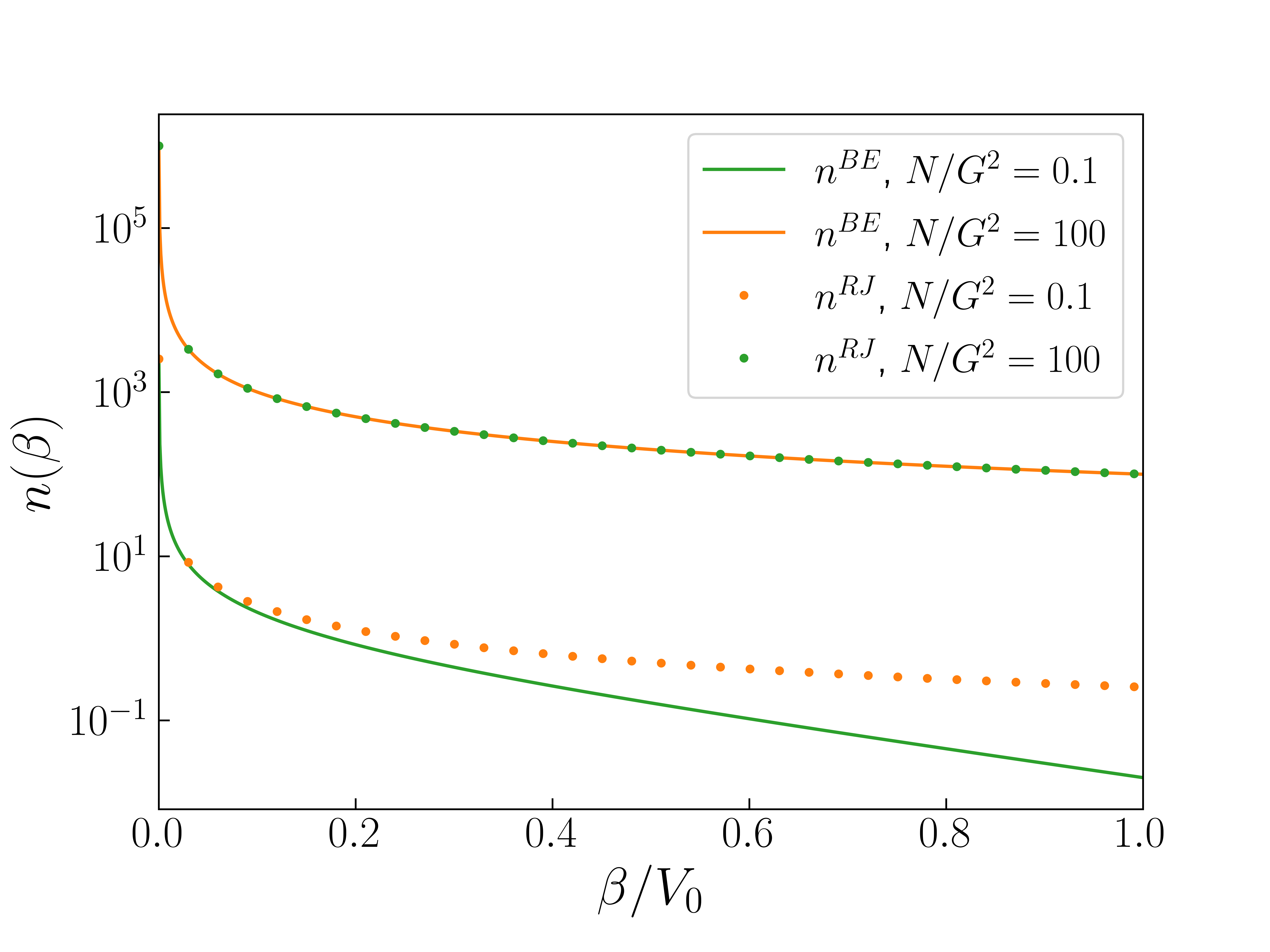}
\caption{
\baselineskip 10pt
{\bf Equilibrium distribution.} 
BE  and RJ equilibrium distributions from Eqs.(\ref{eq:BE_RJ_n_beta}), for $N/G^2=100$ (upper curves), and $N/G^2=0.1$ (lower curves). In the regime $N/G^2 \gg 1$, the RJ distribution $n^{RJ}(\beta)$ (green dots) is a good approximation of the BE distribution $n^{BE}(\beta)$ (solid orange).
For $N/G^2 = 0.1$ (lower curves), the agreement is limited to highly occupied low-order modes, $n^{BE}(\beta) \simeq n^{RJ}(\beta) \gg 1$, whereas a significant discrepancy between the BE (solid green) and RJ (orange dots) distributions emerges for weakly occupied higher-order modes (see the text for details).
}
\label{fig:n_w_BE_RJ} 
\end{figure}

The above picture is consistent with the fact that the RJ distribution closely approximates the exact BE distribution for highly occupied modes.
The BE and RJ distributions are given in Eq.(\ref{eq:BE_RJ_n_beta}).
We report in Fig.~\ref{fig:n_w_BE_RJ} the BE and RJ distributions at ${\tilde T}={\tilde T}_c$, where the value of ${\tilde T}_c$ has been determined by solving the exact relation (\ref{eq:N_int_x}) for a given value of $N/G^2$.
For $N/G^2 \gg 1$, the RJ distribution is a good approximation of the BE distribution. 
Conversely, for $N/G^2=0.1$ the good agreement is limited to highly occupied low-order modes, whereas a deviation between the BE and RJ distributions emerges for weakly populated higher order modes.

\section{Condensate fraction vs energy}
\label{sec:n_0vsE}

As previously discussed  (end of section~\ref{sec:BE_RJ_eq}), in the free propagation geometry considered in  MMFs experiments,  thermalization of the beam of light occurs without any thermostat. 
By keeping $N$ constant,  the control parameter of the transition to condensation is not the temperature, but the energy ${\tilde E}$ of the  speckled beam launched in the MMF, see \cite{PRL20}.
Considering the condensed regime (${\tilde \mu}=\beta_0$), the kinetic energy defined in section~\ref{sec:BE_RJ_eq} then reads 
\begin{align}
{\tilde E} = V_0 G^2 \int_0^1\frac{x^2}{\exp\big( \frac{V_0}{{\tilde T} }x\big) - 1} dx,
\label{eq:E_cond}
\end{align}
and the corresponding critical value of the energy to condensation ${\tilde E}_c$ is obtained by setting ${\tilde T}={\tilde T}_c$ in this equation.
Note that the energy in (\ref{eq:E_cond}) neglects the contribution from the fundamental mode, since as discussed in section~\ref{sec:dos_TL}, in the continuous approximation ${\tilde E} \gg {\tilde E}_0 = N \beta_0$.

Following the analysis in the previous section we obtain in the BE regime 
\begin{align}
\frac{n_0^{BE}}{N} = 1 - \Big(\frac{{\tilde E}}{{\tilde E}_c}\Big)^{\frac{2}{3}},
\label{eq:cond_BE}
\end{align}
with ${\tilde E}_c = 2 \zeta(3) \tilde{T}_c^3/\beta_0^2
 = 2\zeta(3)  N^{3/2}\beta_0/\zeta(2)^{3/2}$.

In the RJ regime, we obtain 
\begin{align}
\frac{n_0^{RJ}}{N} = 1 - \frac{{\tilde E}}{{\tilde E}_c},
\label{eq:cond_RJ}
\end{align}
with ${\tilde E}_c = N V_0/2$. 
The linear behavior of the transition to condensation of classical waves has been reported experimentally in Ref.\cite{PRL20}.
Note in this respect that finite size effects make the transition to condensation smoother, a feature that has been discussed in Ref.\cite{PRL20}.
For clarity, we have reported in Fig.~\ref{fig:n_0vsT_E}(b) the condensate fractions vs energy for the BE and RJ regimes, as well as the intermediate regime for $N/G^2=1$.

\section{Specific heat}

The specific heat is known as an important quantity for characterizing the phase transition of BE condensation \cite{stringari16}.
On the other hand, the properties of the specific heat across the condensation of RJ waves have been discussed in Ref.\cite{PRL20}, while optical calorimetric measurements have been reported in MMFs through the analysis of heat exchanges in Ref.\cite{ferraro24}.

As discussed in the introduction, the specific heat vs temperature has noteworthy distinguished properties for a classical RJ system and a quantum BE system.
In this section we derive a generalized expression of the specific heat that encompasses both the RJ and BE cases, and that recovers the corresponding expressions in their respective limits.
The specific heat is defined by ${\tilde C}_V = (d {\tilde E}/d{\tilde T})_{N,M}$, where the number of modes $M$ plays the role of the `system volume'.
Note that here we have the relation ${\tilde C}_V=C_V/k_B$, where $C_V = (d {E}/d{T})_{N,V}$ is the standard definition of the specific heat.


\subsection{Specific heat in the condensed regime, ${\tilde T} \le {\tilde T}_c$}

\subsubsection{General case}

In the condensed regime ${\tilde T} \le {\tilde T}_c$, we set ${\tilde \mu} = \beta_0$, and by taking the derivative of the energy in Eq.(\ref{eq:E_cond}), we obtain the specific heat:
\begin{align}
{\tilde C}_V = \Big(\frac{V_0}{\tilde T}\Big)^2 G^2 \int_0^1 
\frac{ x^3 \exp(V_0 x /{\tilde T})   }{  \big( \exp(V_0 x /{\tilde T})-1 \big)^2   }
dx.
\label{eq:expressCV}
\end{align}
This expression is general and it is valid for any value of $N/G^2$.
We now discuss the BE and RJ limits.

\subsubsection{BE regime}

In the BE regime  $N \ll G^2$, we have  $V_0/{\tilde T}_c \gg 1$. We also have $V_0/{\tilde T} \gg 1$ since $\tilde{T} \leq \tilde{T}_c$ and we find
\begin{align}
{\tilde C}_V^{BE} = 6 N \frac{\zeta(3)}{\zeta(2)} \Big(\frac{{\tilde T}}{{\tilde T}_c} \Big)^2 .
\label{eq:CV_inf}
\end{align}
This quadratic behavior of the specific heat recovers the expression of a 2D harmonically trapped Bose gas \cite{stringari16}. 
In particular, ${\tilde C}_V^{BE} \to 0$ as $\tilde{T} \to 0$, which is a consequence of the quantum notion of frozen degrees of freedom, see Fig.~\ref{fig:Cv_T}(a).

\subsubsection{RJ regime}

\begin{figure}
\includegraphics[width=.8\columnwidth]{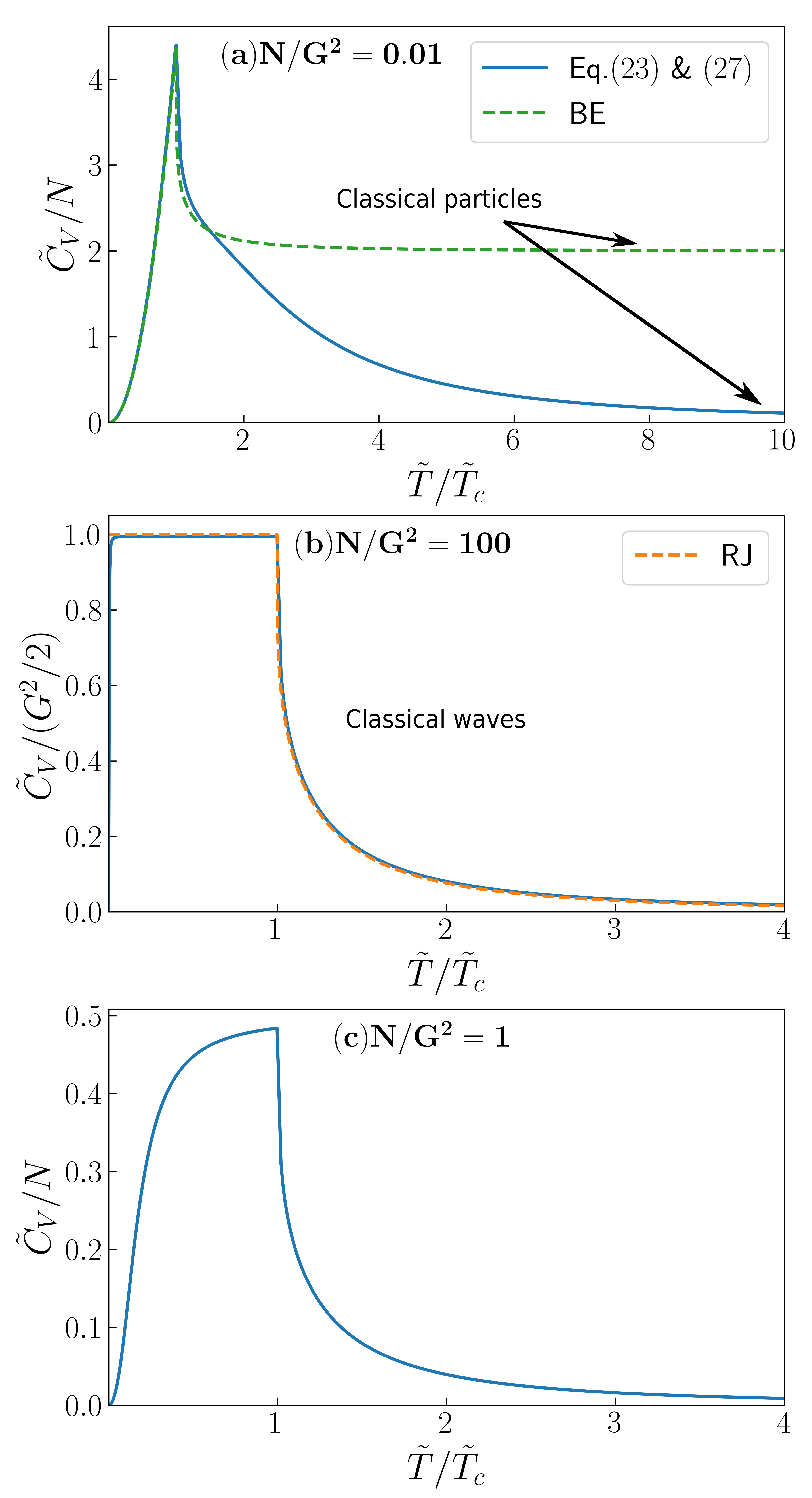}
\caption{
\baselineskip 10pt
{\bf Specific heat for positive temperatures.} 
(a) ${\tilde C}_V ({\tilde T})$ for $N/G^2=0.01$.
The blue line reports the exact general expression [Eq.(\ref{eq:expressCV}) for ${\tilde T} \le {\tilde T}_c$, and Eq.(\ref{eq:CV_gen}) for ${\tilde T} >{\tilde T}_c$].
The dashed green line reports the BE limit [Eq.(\ref{eq:CV_inf}) for ${\tilde T} \le {\tilde T}_c$, and Eq.(\ref{eq:Cv_sup_BE}) for ${\tilde T} >{\tilde T}_c$].
Note the quadratic behavior of ${\tilde C}_V ({\tilde T})$ at small temperatures.
For ${\tilde T}/{\tilde T}_c \gg 1$, the BE case tends to the classical particle limit ${\tilde C}_V^{BE} = 2N$ (energy equipartition among particles), whereas ${\tilde C}_V \to 0$ for the general case because of the finite number of modes of the waveguide.
(b) ${\tilde C}_V ({\tilde T})$ for $N/G^2=100$.
The blue line reports the general expression [Eq.(\ref{eq:expressCV}) for ${\tilde T} \le {\tilde T}_c$, and Eq.(\ref{eq:CV_gen}) for ${\tilde T} >{\tilde T}_c$].
The dashed orange line reports the RJ case [Eq.(\ref{eq:CV_RJ_inf}) for ${\tilde T} \le {\tilde T}_c$, and  Eq.(\ref{eq:c_v_RJ}) for ${\tilde T}>{\tilde T}_c$].
For ${\tilde T} \le {\tilde T}_c$, ${\tilde C}_V^{RJ} = G^2/2$ as a consequence of the classical wave limit (energy equipartition among modes).
(c) ${\tilde C}_V ({\tilde T})$ in the intermediate regime $N/G^2=1$ for the exact general expression [Eq.(\ref{eq:expressCV}) and Eq.(\ref{eq:CV_gen})]: The specific heat exhibits hybrid features inherited from the BE and RJ cases.
}
\label{fig:Cv_T} 
\end{figure}

In the RJ regime $N \gg G^2$, we have $V_0/{\tilde T}_c \ll 1$. As long as we have  $V_0/{\tilde T} \ll 1$, we get  as in \cite{PRL20}:
\begin{align}
{\tilde C}_V^{RJ} = G^2/2 \simeq M.
\label{eq:CV_RJ_inf}
\end{align}
Then we recover the result that, according to the RJ statistics, ${\tilde C}_V^{RJ} \to G^2/2$ for $\tilde{T} \le {\tilde T}_c$, as illustrated in Fig.~\ref{fig:Cv_T}(b).
This behavior is a consequence of the theorem of energy equipartition among the modes inherent to classical statistical mechanics.

\subsubsection{Quantum depletion of the specific heat}

It is interesting to note that  quantum effects can manifest themselves even for $N \gg G^2$.
Indeed, when $\tilde{T}$ becomes small enough so that $V_0/{\tilde T} \sim 1$, then we need to use Eq.(\ref{eq:expressCV}). 
When $\tilde{T}$ is so small  that $V_0/{\tilde T} \gg 1$, then we have ${\tilde C}_V^{BE} = 6 \zeta(3) \tilde{T}^2 / \beta_0^2 = 6 N \zeta(3) (\tilde{T}/\tilde{T}_c)^2 (N/G^2)$.
In this regime  ${\tilde E}/{\tilde E}_0=2 \zeta(3)({\tilde T}/V_0)^3 (G^2/N)  G$, where ${\tilde E}_0=N \beta_0$ denotes the lowest energy when all particles populate the fundamental mode.
Accordingly, an appreciable deviation from the RJ behavior occurs for $G \gg N/G^2 \gg 1$.
This is illustrated in Fig.~\ref{fig:Cv_deplet}, 
which evidences the quantum depletion of the specific heat over a large number ($\sim 30$) of groups of modes -- note that ${\tilde C}_V$ deviates from the RJ value $G^2/2$ for ${\tilde E} \lesssim 30 {\tilde E}_0$.
Such unexpected manifestation of quantum effects for $N \gg G^2$ is due to the strong condensate fraction for ${\tilde E} \lesssim 30 {\tilde E}_0$ (typically larger than 80\% in Fig.~\ref{fig:Cv_deplet}), so that the RJ condition $(N-n_0)/G^2 \gg 1$ is no longer verified. 
In other terms, the population of the modes gets rarefied and the condition $n_{p \neq 0} \gg 1$ is no longer verified, which invalidates the RJ statistics.

\subsection{Specific heat in the normal regime, ${\tilde T} > {\tilde T}_c$}

\subsubsection{General case}

Above the transition to condensation ${\tilde T} > {\tilde T}_c$, the chemical potential verifies ${\tilde \mu} < \beta_0$, as discussed above through Fig.~\ref{fig:0}(a).
The expression of the energy takes the general form
\begin{align}
{\tilde E} = V_0 G^2 \int_0^1\frac{x^2}{z^{-1}\exp\big( \frac{V_0}{{\tilde T} }x\big) - 1} dx,
\label{eq:E_z}
\end{align}
where we recall that $z=\exp({\hat \mu}/{\tilde T})$ is the fugacity, with ${\hat \mu} = {\tilde \mu}-\beta_0$, which is a function of the temperature, ${\hat \mu}({\tilde T})$.
Since ${\tilde T} > {\tilde T}_c$, we have $z < 1$, see Fig.~\ref{fig:0}(b).

\begin{figure}
\includegraphics[width=.8\columnwidth]{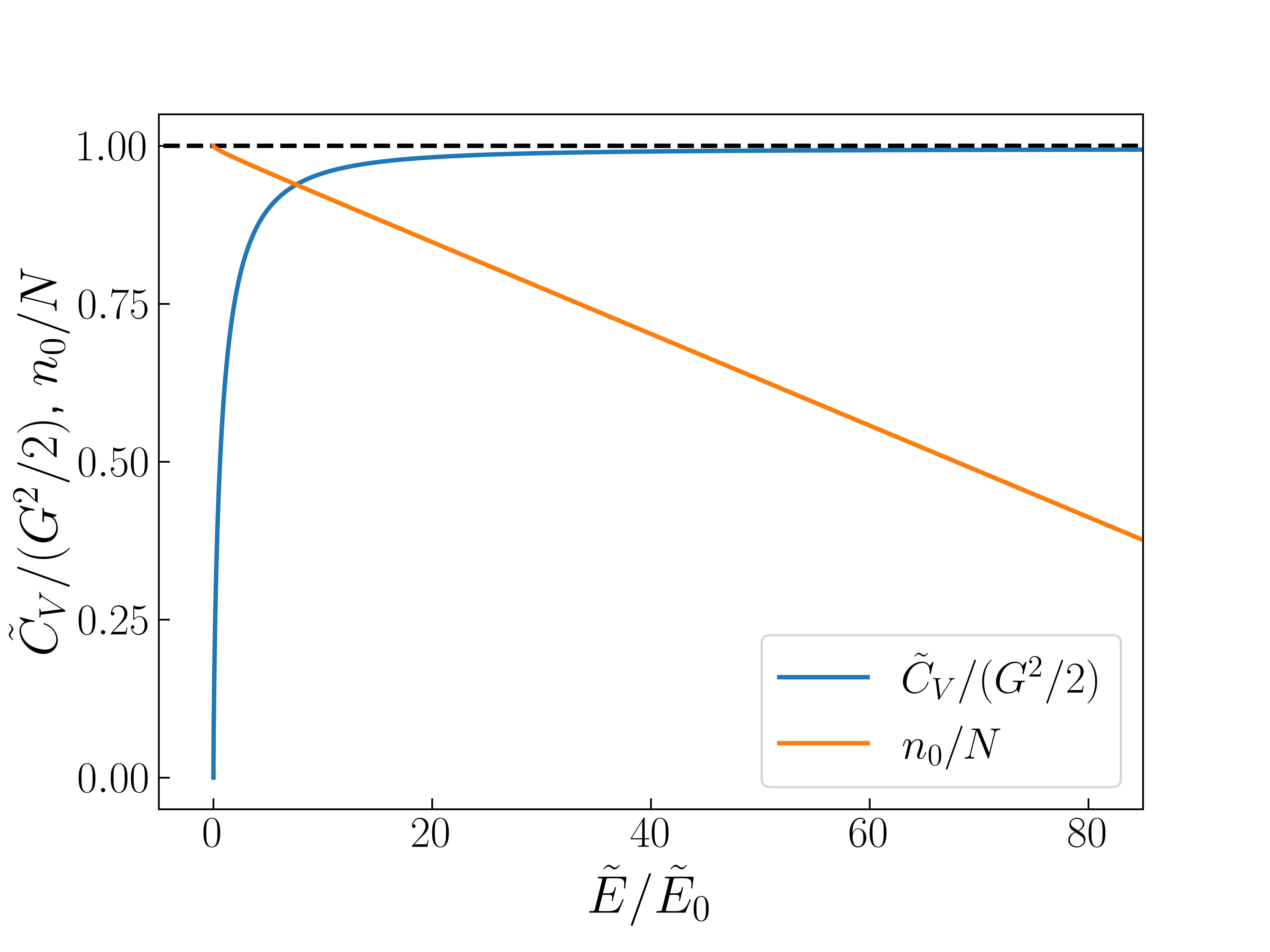}
\caption{
\baselineskip 10pt
{\bf Quantum depletion of the specific heat for $N/G^2 \gg 1$.} 
Specific heat ${\tilde C}_V$ vs ${\tilde E}/{\tilde E}_0$ [from Eq.(\ref{eq:expressCV})] for $N/G^2=10$ 
(blue line).
Corresponding condensate fraction $n_0/N$ vs ${\tilde E}/{\tilde E}_0$ (orange line).
Note the depletion of ${\tilde C}_V \to 0$ for ${\tilde E} \lesssim 30 {\tilde E}_0$, i.e., over $\sim 30$ groups of modes.
This manifestation of quantum effects for $N/G^2 \gg 1$ is due to the strong population of the fundamental mode ($\gtrsim 80$\%), which rarefies the population of the other modes and thus invalidates the RJ approximation.
}
\label{fig:Cv_deplet} 
\end{figure}

By computing the derivative of the energy Eq.(\ref{eq:E_z}) with respect to the temperature, we obtain
\begin{align}
{\tilde C}_V({\tilde T}) = \frac{V_0}{\tilde T} G^2 
\int_0^{1} 
&\frac{x^2 z^{-1}\exp\big( \frac{V_0}{\tilde T}x \big) }{ \big[z^{-1} \exp\big( \frac{V_0}{\tilde T}x \big) - 1  \big]^2  } 
\nonumber \\  
& \times \ \left[ \frac{V_0 }{{\tilde T}}x +\frac{{\tilde T}}{z}\frac{dz}{d{\tilde T}}    \right] dx.
\label{eq:CV_gen}
\end{align}
In this expression, we need the function $z(\tilde{T})$.
As discussed in section~\ref{sec:converg_TL}, this function is obtained by numerically solving Eq.(\ref{eq:N_int_x_basic})
for $V_0$ and $N/G^2$ given.
Typical examples of $z(\tilde{T})$ are reported in panels (b) of Figs.~\ref{fig:0}-\ref{fig:0_BE} in the RJ, the BE, and the intermediate regimes, which show that $z$ reaches $1$ at ${\tilde T} = {\tilde T}_c$, as expected.
Furthermore the specific heat is computed by keeping constant $N$, so that  $dN/d{\tilde T}=0$ gives
\begin{align}
\frac{{\tilde T}}{z}\frac{dz}{d{\tilde T}}
=
-\frac{V_0}{{\tilde T}}
\frac{  \int_0^{1} \frac{x^2 \exp(V_0 x/{\tilde T}) }{[z^{-1}\exp( V_0x/{\tilde T}  ) - 1]^2} dx  }{   \int_0^{1} \frac{x \exp(V_0 x/{\tilde T})  }{[z^{-1}\exp( V_0x/{\tilde T}  ) - 1]^2}   dx  }.
\label{eq:TdmusdT}
\end{align}
Once $z(\tilde{T})$  is obtained by solving (\ref{eq:N_int_x_basic}), it can be substituted in 
Eq.(\ref{eq:TdmusdT}).
The corresponding expression of $\frac{{\tilde T}}{z}\frac{dz}{d{\tilde T}}$ is then substituted in Eq.(\ref{eq:CV_gen}) to get ${\tilde C}_V({\tilde T})$.
We report in Fig.~\ref{fig:Cv_T}(a) and (b) the specific heat $C_V({\tilde T})$ in the BE and RJ regimes, respectively, as well as in the intermediate regime $N/G^2=1$ in panel \ref{fig:Cv_T}(c), where it exhibits properties reminiscent of both the BE and RJ regimes as we will now discuss in more detail.

\subsubsection{BE regime}

In the BE regime we have $V_0/\tilde{T}_c \gg 1$.
For $\tilde{T}> \tilde{T}_c$ so that $V_0/\tilde{T} \gg 1$, we get
\begin{align}
{\tilde E} = V_0 G^2 \int_0^\infty \frac{x^2}{z^{-1}\exp\big( \frac{V_0}{{\tilde T} }x\big) - 1} dx = 2\frac{{\tilde T}^3}{\beta_0^2} g_3(z) ,
\end{align}
where $g_p(z)=\frac{1}{\Gamma(p)}\int_0^\infty dx \frac{x^{p-1}}{z^{-1} e^x-1}=\sum_{l=1}^\infty \frac{z^l}{l^p}$ is the Bose function \cite{stringari16}. 
By the property $g_p'(z)=z^{-1}g_{p-1}(z)$, we find
\begin{align}
{\tilde C}_V^{BE} = \frac{{\tilde T}^2}{\beta_0^2}
\Big( 6 g_3(z) + 2  \frac{\tilde{T}}{z}\frac{dz}{d{\tilde T}} g_2(z) \Big).
\end{align}
In the BE regime Eqs.(\ref{eq:N_int_x_basic}) and (\ref{eq:TdmusdT}) can be simplified by letting the upper bounds of the integrals go to infinity:
\begin{align}
N = \frac{\tilde{T}^2}{\beta_0^2} g_2(z),\qquad 
\frac{\tilde T}{z}\frac{dz}{d{\tilde T}}=-2\frac{g_2(z)}{g_1(z)},
\end{align}
and we obtain
\begin{align}
{\tilde C}_V^{BE} = 2N \Big( 3 \frac{g_3(z)}{g_2(z)}- 2 \frac{g_2(z)}{g_1(z)} \Big),
\label{eq:Cv_sup_BE}
\end{align}
where $z$  can be determined from $\tilde{T}$ by $g_2(z) = ({\tilde{T}_c}/{\tilde{T}})^2 \zeta(2)$.

As discussed above through Fig.~\ref{fig:0}(a-b), by decreasing the temperature ${\tilde T} \to {\tilde T}_c^+$, the chemical potential increases  ${\tilde \mu} \to \beta_0^-$, and the fugacity increases to one, $z\to 1^-$.
Recalling that $g_1(z)=-\log(1-z)$, then $\zeta(1)=\infty$, which gives
${\tilde C}_V^{BE} =6N\zeta(3)/\zeta(2)$.
On the other hand, according to Eq.(\ref{eq:CV_inf}), for ${\tilde T} \to {\tilde T}_c^-$, ${\tilde C}_V^{BE}=6N\zeta(3)/\zeta(2)$.
Then the specific heat is continuous at the transition to condensation, which is in contrast with the case of a 3D parabolic potential, where the specific heat is known to exhibit a discontinuity \cite{stringari16}. 
Note however that the specific heat exhibits a pronounced cusp at ${\tilde T}_c$, see Fig.~\ref{fig:Cv_T}(a).

For ${\tilde T} \gg {\tilde T}_c$, $z \to 0$ and $g_p(z)\simeq z$, so that ${\tilde C}_V^{BE}=2N$, see Fig.~\ref{fig:Cv_T}(a).
Recalling that ${\tilde C}_V=C_V/k_B$, we obtain $C_V^{BE}/(N k_B)=2$.
Indeed, at high temperatures, the Bose gas behaves as a classical ideal gas, and the equipartition theorem of classical statistical mechanics stipulates that each quadratic degree of freedom contribute $k_B T/2$ to the average value of the energy.
Here, the quadratic contribution of the kinetic and potential energy add each other in 2D for $N$ particles, hence $E=2N k_B T$. 
Note that this result, which is inherent to the classical limit of particles (Boltzmann equilibrium distribution), differs from the {classical limit of waves} (RJ equilibrium distribution), discussed above through the RJ distribution, see Eq.(\ref{eq:CV_RJ_inf}).

\begin{figure}
\includegraphics[width=.8\columnwidth]{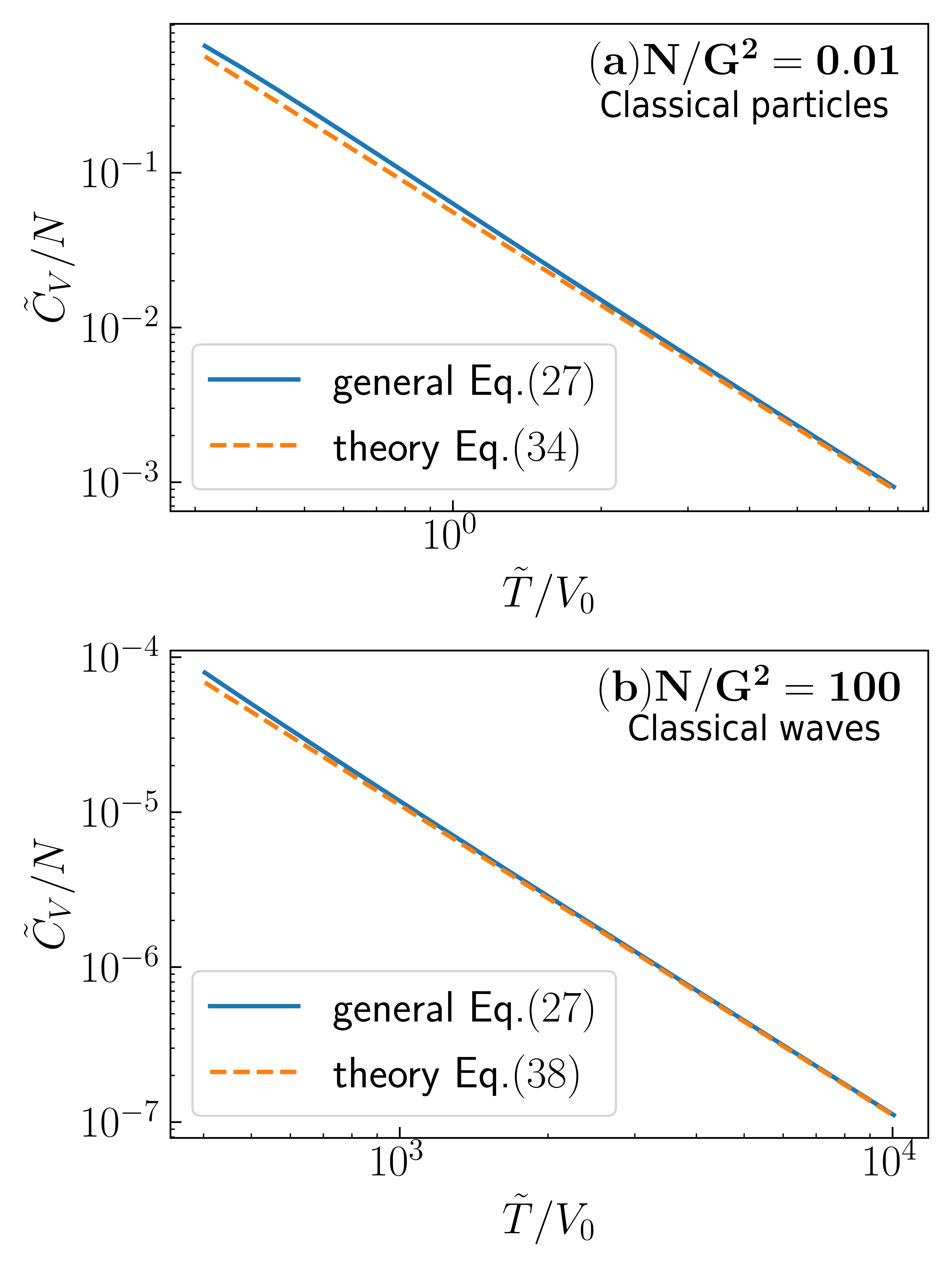}
\caption{
\baselineskip 10pt
{\bf Large temperature behavior of ${\tilde C}_V \sim 1/{\tilde T}^2$.} 
Because of the finite number of modes inherent to the guided configuration of the fluid of light, the specific heat tends to zero at large temperatures.
The blue line reports ${\tilde C}_V/N$ vs  ${\tilde T}/V_0$ from the general expression in Eq.(\ref{eq:CV_gen}): For the classical (Boltzmann) particle limit $N/G^2 = 0.01$ (a); for the classical  RJ wave limit  $N/G^2 = 100$ (b).
The orange line reports the corresponding theoretical predictions: Eq.(\ref{eq:Cv_B_Tlarge}) for (a);  Eq.(\ref{eq:C_V_RJ_Tlarge}) for (b).
The log-log plot evidences the predicted power-law decay,  ${\tilde C}_V \sim 1/{\tilde T}^2$.
}
\label{fig:loglogCV} 
\end{figure}

The energy equipartition, and the fact that the specific heat takes a constant value ${\tilde C}_V^{BE}=2N$ for ${\tilde T} \gg {\tilde T}_c$ is in marked contrast with the decay ${\tilde C}_V^{BE} \to 0$ evidenced in Fig.~\ref{fig:Cv_T}(a) for the waveguide configuration (blue line).
We derive in the Appendix the large temperature asymptotic behavior of the fugacity and the specific heat for the BE case in the classical (Boltzmann) gas limit in the presence of a frequency cut-off:
\begin{align}
&z \simeq \frac{2 N}{G^2} + \frac{4 N {V_0}}{3 G^2 \tilde T} , 
\label{eq:z_B_Tlarge}\\
&{\tilde C}_V^{BE} \simeq    \frac{N V_0^2}{18 \tilde T^2} .
\label{eq:Cv_B_Tlarge}
\end{align}
The quadratic algebraic decay of the specific heat for large temperatures in Eq.(\ref{eq:Cv_B_Tlarge}) is confirmed by the plot of the exact general expression of ${\tilde C}_V^{BE}({\tilde T})$ given in Eq.(\ref{eq:CV_gen}), as evidenced in Fig.~\ref{fig:loglogCV}(a).

\subsubsection{RJ regime}

In the RJ regime we have $V_0/\tilde{T}_c \ll 1$.
For all $\tilde{T}> \tilde{T}_c$ we have $V_0/\tilde{T} \ll 1$.
Considering the RJ limit of Eq.(\ref{eq:CV_gen}) and Eq.(\ref{eq:N_int_x_basic}), we obtain after a lengthy calculation:
\begin{eqnarray}
\frac{{\tilde C}_V^{RJ}({\tilde \mu})}{G^2/2}= 1 + \frac{2[ 1+({\hat \mu}/V_0) \log((V_0-{\hat \mu})/(-{\hat \mu})  ]^2}
{ V_0/(V_0-{\hat \mu}) - \log\big((V_0-{\hat \mu})/(-{\hat \mu}) \big)  }
\label{eq:c_v_RJ}
\end{eqnarray}
where we recall that ${\hat \mu}={\tilde \mu}-\beta_0$.
This expression of the specific heat recovers the expression obtained in Ref.\cite{PRL20} (see Eq.(15) of the Supplementary Material), with the number of modes $M \simeq G^2/2$. 
To obtain the specific heat as a function of the temperature ${\tilde C}_V^{RJ}({\tilde T})$, we make use of $N={\tilde T}\sum_p^G 1/(\beta_p-\mu)$, that reads in the continuous limit:
\begin{eqnarray}
\frac{{\tilde T}({\hat \mu})}{{\tilde E}_0}= 
\frac{1}{G+({\hat \mu}/\beta_0)  \log(1-V_0/{\hat \mu}) },
\label{eq:TsNbeta0_RJ}
\end{eqnarray}
where we recall that ${\tilde E}_0=N\beta_0$.
The parametric plot of Eq.(\ref{eq:c_v_RJ}) and Eq.(\ref{eq:TsNbeta0_RJ}) with respect to ${\hat \mu}$ gives ${\tilde C}_V^{RJ}({\tilde T})$, see Fig.~\ref{fig:Cv_T}(b).
Note in Fig.~\ref{fig:Cv_T}(b) that $C_V^{RJ}$ exhibits a cusp at ${\tilde T}={\tilde T}_c$:
\begin{eqnarray}
\lim\limits_{
\begin{array}{l}
{\tilde T} \to {\tilde T}_{c}^+
\end{array}}
\frac{\partial {\tilde C}_V^{RJ}}{\partial {\tilde T}} \ = \!
- \infty.
\label{eq:dC_VsurdT}
\end{eqnarray}

Let us now consider the specific heat at large temperatures.  
The expansion of the general expression Eq.(\ref{eq:c_v_RJ}) provides the following asymptotic behavior for ${\tilde T} \gg {\tilde T}_c$:
\begin{eqnarray}
{\tilde C}_V^{RJ} \simeq \frac{1}{9}\frac{N^2 \beta_0^2}{{\tilde T}^2}. 
\label{eq:C_V_RJ_Tlarge}
\end{eqnarray}
This algebraic decay of the specific heat for large temperatures is confirmed by the behavior of the general expression of ${\tilde C}_V({\tilde T})$ given in Eq.(\ref{eq:CV_gen}), as illustrated in Fig.~\ref{fig:loglogCV}(a).
It is interesting to note that this algebraic scaling is the same as that found above in the regime $N/G^2 \ll 1$, see Eq.(\ref{eq:Cv_B_Tlarge}).

We conclude this section by recalling that the decay to zero of the specific heat at large temperatures, ${\tilde C}_V \to 0$ for ${\tilde T} \to \infty$, is a consequence of the finite number of modes inherent to the optical waveguide configuration. 
This aspect will be discussed in more detail in the forthcoming section devoted to negative temperatures.



\begin{figure}
\includegraphics[width=.8\columnwidth]{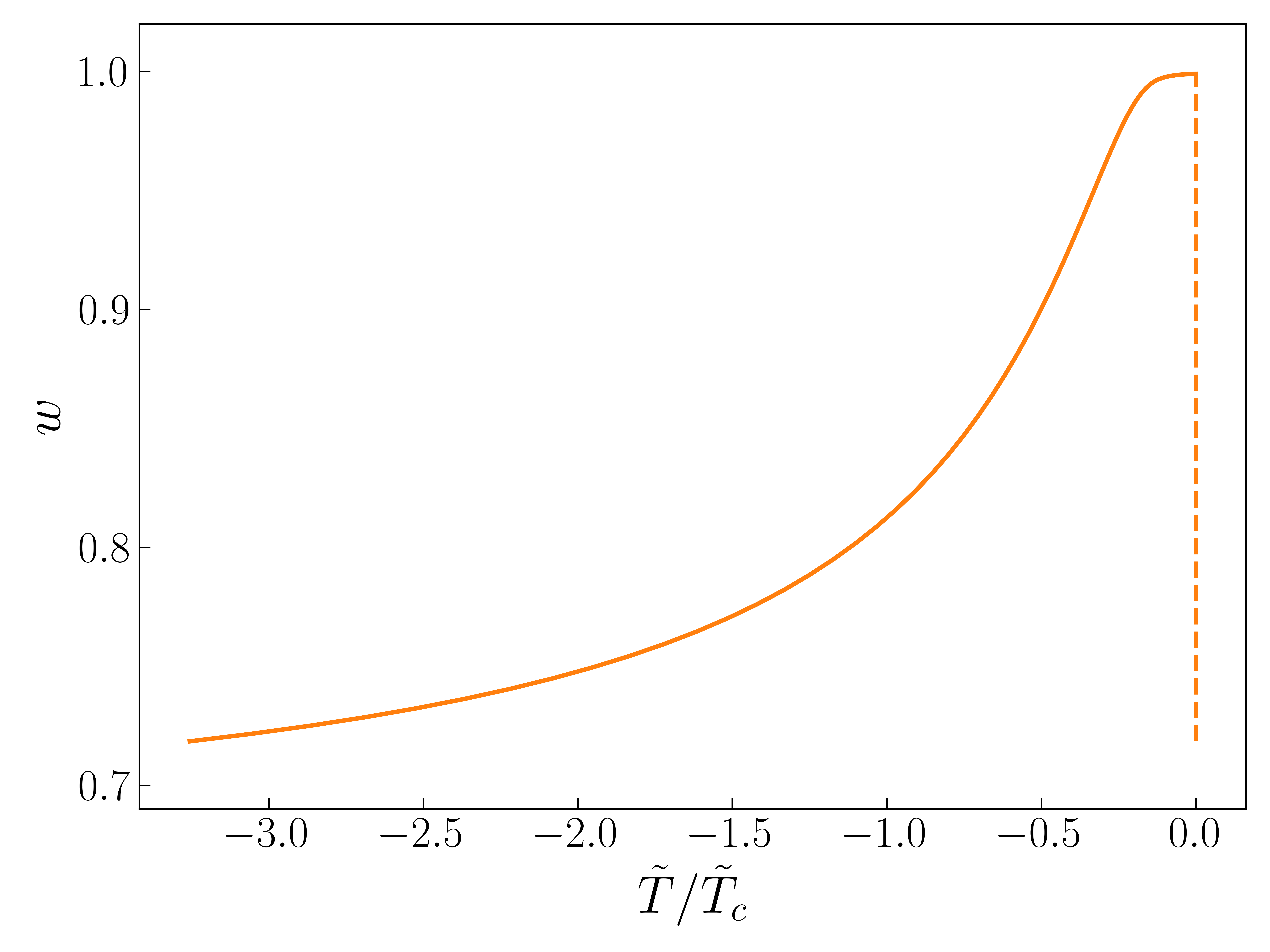}
\caption{
\baselineskip 10pt
{\bf Fugacity at negative temperatures.} 
Fugacity $w=e^{({\tilde \mu}-V_0)/{\tilde T}}$ at negative temperatures ${\tilde T}<0$ obtained by solving Eq.(\ref{eq:N_Tneg}).
Note that $w \to 1^-$ only at zero temperature: The inverted condensation of particles in the highest energy level does not occur in the thermodynamic limit (compare to Fig.~\ref{fig:0}(b) where $z \to 1^-$ at ${\tilde T}={\tilde T}_c > 0$, as expected from the phase transition to condensation for positive temperatures).
}
\label{fig:z_w} 
\end{figure}

\section{Negative temperatures for the BE statistics}

\begin{figure}
\includegraphics[width=.95\columnwidth]{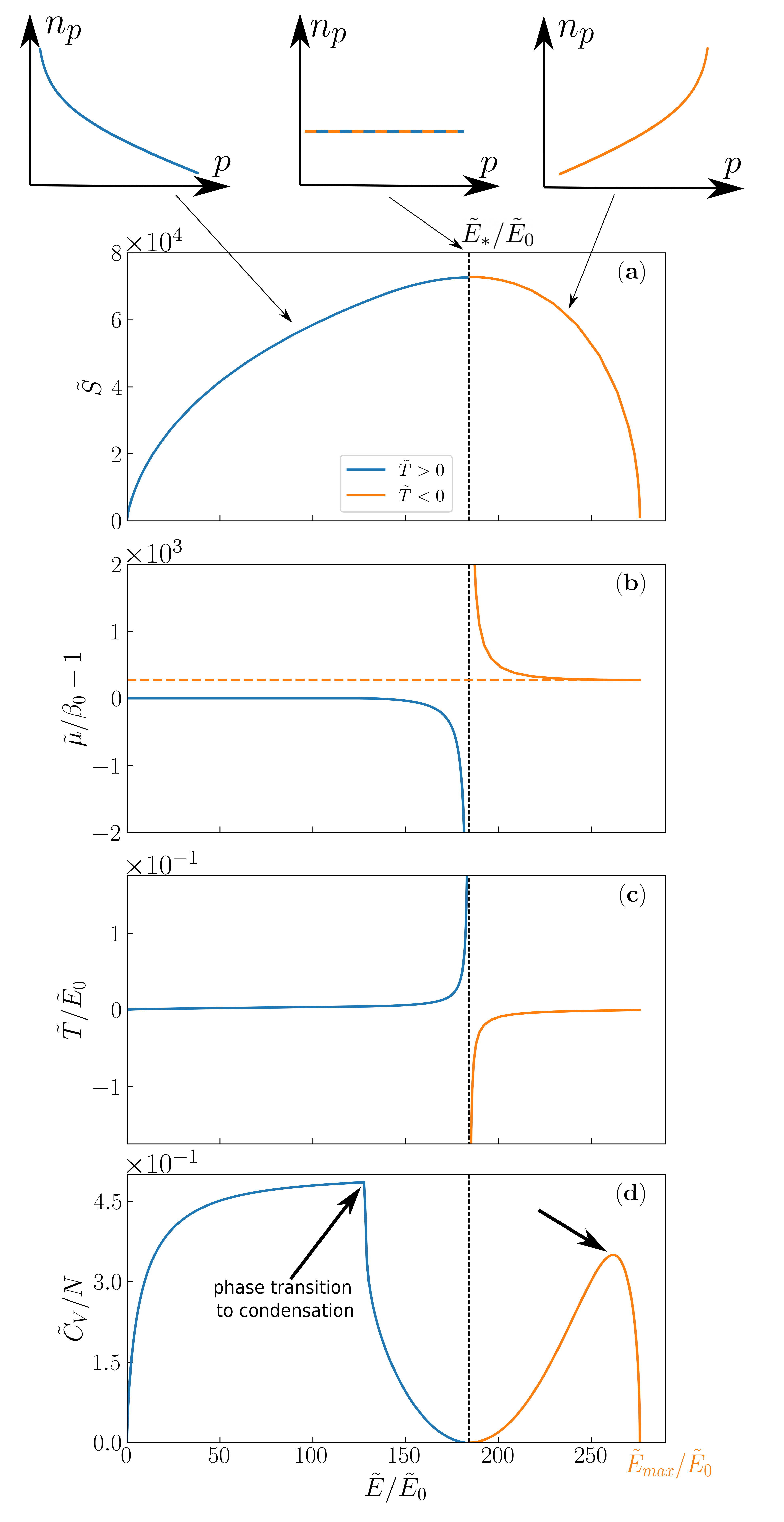}
\caption{
\baselineskip 10pt
{\bf Negative temperatures.} 
(a) Entropy ${\tilde S}$ vs energy: 
The vertical dashed line at ${\tilde E}={\tilde E}_*$ separates positive and negative temperatures.
The upper panels report equilibrium distributions for positive temperatures ${\tilde T}>0$ (${\tilde E}<{\tilde E}_*$), 
and negative temperatures ${\tilde T}<0$ (${\tilde E}>{\tilde E}_*$) featured by an inverted population, while for $1/{\tilde T} \to 0$ (${\tilde E}={\tilde E}_*$),   the equilibrium distribution is homogeneous.
(b) Chemical potential ${\tilde \mu}/\beta_0 - 1$ vs energy: 
Note the asymptotic behavior ${\tilde \mu} \to \beta_0^-$ for ${\tilde E} \to 0$ (horizontal dashed line) corresponding to condensation in the lowest energy level (fundamental mode), and ${\tilde \mu} \to V_0^+$ for ${\tilde E} \to {\tilde E}_{\rm max}=G {\tilde E}_0$ corresponding to a macroscopic population of the highest energy level (highest-order
mode group).
(c) Temperature vs energy. 
(d) ${\tilde C}_V$ vs ${\tilde E}/{\tilde E}_0$ for $N/G^2=1$. 
Note that ${\tilde C}_V \to 0$ for ${\tilde T} \to 0^+$ (i.e., ${\tilde E} \to 0$), and for ${\tilde T} \to 0^-$ (i.e., ${\tilde E} \to {\tilde E}_{\rm max}$).
${\tilde C}_V$ exhibits a cusp at the transition to condensation ${\tilde T}={\tilde T}_c>0$.
However, the cusp gets smoothed at ${\tilde T}<0$ (see the right arrow in (d)), because the inverted condensation leading to the macroscopic population of the highest energy level does not occur in the thermodynamic limit.
}
\label{fig:Cv_E} 
\end{figure}

The physical idea of negative temperatures was originally conceived in the seminal works by Onsager \cite{onsager49} and Ramsey \cite{ramsey56}. 
During the last decades, many works have been devoted to the theoretical understanding of these unusual equilibrium states. 
Negative temperatures are now broadly accepted, in line with different experimental observations \cite{frenkel15,buonsante16,puglisi17,cerino15,baldovin21,onorato21}. 
As a matter of fact, negative temperatures were originally observed experimentally in nuclear spin systems \cite{purcell51}, and more recently with cold atoms in optical lattices \cite{braun13}.
Furthermore, negative temperatures originally predicted by Onsager in the statistical description of point vortices \cite{onsager49} have been recently observed in 2D quantum superfluids \cite{gauthier19,johnstone19}. 
More recently, negative temperatures have been predicted for classical multimode optical wave systems in the seminal work \cite{christodoulides19}, and subsequently observed in these systems with light waves in Refs.~\cite{PRL23,muniz23}.
In particular, Ref.\cite{PRL23} has reported the observation of light thermalization to negative temperatures RJ states in the free propagation of speckled beams in MMFs.
In the following we extend the existence of such negative temperatures equilibria to the quantum regime, see Fig.~\ref{fig:Cv_E}.

\subsection{Energy and entropy for negative temperatures}

We recall that negative temperatures owe their existence to the finite number of modes supported by the fiber.
As a consequence, the energy spectrum is bounded from above 
${\tilde E} \le E_{\rm max}=G {\tilde E}_0$,
where $E_{\max}$ is the highest value of the energy, obtained when all photons populate the highest energy level.
Such a macroscopic population of the highest energy level may suggest an inverted condensation to occur at negative temperatures. 
To analyze this aspect, we first note that for ${\tilde T}<0$, the condition $n_p^{BE} \ge 0$ imposes ${\tilde \mu} \ge V_0$, see the horizontal dashed line in Fig.~\ref{fig:Cv_E}(b).
Accordingly, the number of photons can be appropriately written for negative temperatures in the following form
\begin{align}
\frac{N}{G^2} =\int_0^1 \frac{1-x}{w^{-1}\exp\big( \frac{V_0 }{-{\tilde T}}x  \big) - 1} dx,
\label{eq:N_Tneg}
\end{align}
where we have introduced an analogous of the fugacity for negative temperatures
\begin{align}
0<w=e^{({\tilde \mu}-V_0)/{\tilde T}}\le 1.
\label{eq:w}
\end{align}
Indeed, in analogy with positive temperatures where $z \to 1^-$ for ${\tilde T} \to 0^+$, for negative temperatures we have
$$
w \to 1^- \quad {\rm for} \quad {\tilde T} \to 0^-,
$$ 
which leads to a macroscopic population of the highest mode group at the energy level ${\tilde E}_{\rm max}$.
However, at variance with BE condensation in the fundamental mode for ${\tilde T}>0$, negative temperature condensation in the highest energy level does not occur in the thermodynamic limit because the integral 
\begin{align}
\frac{N}{G^2} =\int_0^1 \frac{1-x}{\exp\big( \frac{V_0 }{-{\tilde T}_c}x  \big) - 1} dx
\label{eq:N_Tneg_w1}
\end{align}
does not converge for $x = 0$.
This means that ${\tilde \mu}$ does not reach $V_0$ for a non-vanishing critical temperature ${\tilde T}_c < 0$.
This is illustrated in Fig.~\ref{fig:z_w}(b), which shows that $w \to 1$ only at zero temperature.

The expression of the beam energy is obtained, in a similar way, as 
\begin{align}
\frac{\tilde E}{{\tilde E}_0} =\frac{G^3}{N} \int_0^1 \frac{(1-x)^2}{w^{-1}\exp\big( \frac{V_0 }{-{\tilde T}}x  \big) - 1} dx.
\label{eq:E_Tneg}
\end{align}
For a given value of $N/G^2$, Eq.(\ref{eq:N_Tneg}) can be solved to determine the function $w({\tilde T})$, which can be substituted in (\ref{eq:E_Tneg}) to get ${\tilde E}({\tilde T})$, see Fig.~\ref{fig:Cv_E}(c).

Let us now analyze the evolution of the entropy with the energy, where we recall the basic definition of the temperature, $1/T = (\partial S/\partial E)_{M,N}$. 
The entropy is given by ${\tilde S}=S/k_B=\sum_p^G (n_p^{BE}+1)\log(n_p^{BE}+1)-n_p^{BE}\log(n_p^{BE})$ \cite{stringari16}, which can also be written in the form:
\begin{align*}
{\tilde S}=\frac{1}{{\tilde T}}\sum_p^G \frac{\beta_p-{\tilde \mu}}{e^{(\beta_p-{\tilde \mu})/{\tilde T}}-1} 
- \sum_p^G \log\big(1-e^{({\tilde \mu}-\beta_p)/{\tilde T}}\big).
\end{align*}
By taking the continuous limit, we obtain the expression of the entropy
\begin{align}
{\tilde S} =  \, &G^2
\int_0^1 \frac{(1-x)\big[V_0 x/(-{\tilde T})- \log(w)\big]}{w^{-1}\exp\big( \frac{V_0 x}{-{\tilde T}} \big) - 1} dx
\nonumber\\
&- G^2 \int_0^1 (1-x) \log\big[ 1-w \exp(V_0 x/{\tilde T})\big]dx.
\label{eq:S_gen}
\end{align}

The evolution of the entropy with the energy is reported in Fig.~\ref{fig:Cv_E}(a).
Starting at minimum energy where only the fundamental mode is populated, an increase in energy leads to an occupation of a larger number of fiber modes and therefore an increase of entropy. 
As the temperature approaches infinity, all fiber modes become equally populated, see the first row in Fig.~\ref{fig:Cv_E}.
This refers to the most disordered state where the entropy reaches a maximum for $n_p^{BE}=n_*$, so that $N=n_* G^2/2$ and ${\tilde E}_*={\tilde E}_0(1+2G/3)$. 
Negative temperatures equilibrium states occur for $E > E_*$, where the entropy decreases by increasing the energy.
In this case, higher-order modes are more populated than low-order modes, which leads to an inverted modal population, see the first row of Fig.~\ref{fig:Cv_E}. 


\subsection{Heat capacity for negative temperatures}

The specific heat for negative temperatures is obtained by following a procedure similar to that of positive temperatures. 
We obtain
\begin{align}
{\tilde C}_V = \frac{V_0}{\tilde T} G^2 
\int_0^{1} 
&\frac{(1-x)^2 w^{-1}\exp\big( \frac{V_0}{-\tilde T}x \big) }{ \big[w^{-1} \exp\big( \frac{V_0}{-\tilde T}x \big) - 1  \big]^2  } 
\nonumber \\  
& \times \ \left[ \frac{V_0 }{-{\tilde T}}x +\frac{{\tilde T}}{w}\frac{dw}{d{\tilde T}}    \right] dx.
\label{eq:CV_gen_Tneg}
\end{align}
The parametric plot of ${\tilde C}_V$ in Eq.(\ref{eq:CV_gen_Tneg}) and ${\tilde E}$ in Eq.(\ref{eq:E_Tneg}) with respect to the temperature provides ${\tilde C}_V$ vs ${\tilde E}/{\tilde E}_0$, which is reported in Fig.~\ref{fig:Cv_E}(d).

We have already discussed in detail  the properties of the specific heat for positive temperatures through Fig.~\ref{fig:Cv_T}.
Nearby the transition to negative temperatures, we note that ${\tilde C}_V \to 0$ for ${\tilde E} \to {\tilde E}_*$ \cite{PRL23}. 
This is because the temperature diverges to infinity, ${\tilde T} \to \pm \infty$ when ${\tilde E} \to {\tilde E}_*^{\mp}$, so that a variation of ${\tilde T}$ does not significantly affect ${\tilde E}$, which entails ${\tilde C}_V  \to 0$ nearby ${\tilde E}_*$.
This means that the equilibrium is not constrained by the conservation of the energy  (the Lagrange multiplier $1/{\tilde T}$ is zero), but solely by the conservation of $N$, which merely explains the homogeneous nature of the modal distribution for ${\tilde E}={\tilde E}_*$ in Fig.~\ref{fig:Cv_E} (first row), see Ref.\cite{PRX17}.

Next, we note that the specific heat tends to zero for negative temperatures when the highest energy level becomes macroscopically populated, ${\tilde C}_V \to 0$ for ${\tilde T} \to 0^-$.
This behavior is reminiscent of the quantum property of frozen degrees of freedom, in a way similar to positive temperatures when the fundamental mode becomes macroscopically populated, ${\tilde C}_V \to 0$ for ${\tilde T} \to 0^+$.
Note however a fundamental difference between positive and negative temperatures: 
For positive temperatures the specific heat exhibits a cusp at the critical energy ${\tilde E}_c$ (or temperature ${\tilde T}_c$ in Fig.~\ref{fig:Cv_T}) of the transition to condensation.
In contrast, for negative temperatures, the cusp of the specific heat is smoothed out, see the arrow in Fig.~\ref{fig:Cv_E}(d).
This crucial difference is due to the fact that the transition to condensation is a genuine phase transition for positive temperatures, whereas for negative temperatures there is no phase transition, as discussed above through Eq.(\ref{eq:N_Tneg_w1}).
In other words, the cusp (smooth) behavior of the specific heat for positive (negative) temperatures reflects the critical (regular) nature of the transition toward a macroscopic population of the lowest (highest) energy level.


\section{Discussion and conclusion}

We have clarified the links between the effect of RJ condensation recently observed in MMFs, and the well documented equilibrium properties of the quantum BE condensation. 
By introducing a frequency cut-off inherent to the guided propagation of a beam of light, we have derived generalized expressions of the condensate fraction and the specific heat, which include the previously known RJ and BE limits as particular cases.
In this way, we have described the crossover between classical RJ waves and quantum BE particles.
In particular, we have reconciled the quantum property of frozen degrees of freedom, characterized by a vanishing specific heat at zero temperatures (${C}_V \to 0$ as ${T} \to 0$), with the classical RJ wave limit exhibiting constant specific heat (${\tilde C}_V=C_V/k_B \to M$ for $T \ll {T}_c$). 
We have then shown that the quantum depletion of the specific heat at small temperatures can also occur in the supposedly  classical regime ($N \gg M$) in the presence of a strong condensate, see Fig.~\ref{fig:Cv_deplet}.  
Our analysis also distinguishes the classical RJ wave limit characterized by an energy equipartition among the modes at small temperatures, from the classical particle limit at high temperatures that is characterized by an energy equipartition among the particles.
In addition, we have shown that the specific heat exhibits a universal algebraic decay law $C_V \sim 1/T^2$ for $T \to \infty$, for both the  classical particle (Boltzmann) limit ($N \ll M$) and the classical RJ wave limit ($N \gg M$), see Fig.~\ref{fig:loglogCV}.
Furthermore, the analysis of the specific heat at negative temperatures exhibits distinctive properties, when compared with the case of positive temperatures, see Fig.~\ref{fig:Cv_E}(d).
In particular, the cusp of the specific heat characterizing the phase transition to condensation for positive critical temperature ($T_c >0$) is smoothed out at negative temperatures, reflecting the non-critical nature of the transition to a macroscopic population of the highest energy level.

We recall that, in our study, we have considered the case of a parabolic trapping potential.
The parabolic potential is a significant representative example, because it corresponds to the experiments in MMFs where RJ condensation has been observed \cite{PRL20,mangini22,pourbeyram22,ferraro23}.
Our analysis can be extended to other forms of potentials.
The most natural one could be the homogeneous trapping potential, corresponding to the so-called step-index MMFs.
However, the density of states is constant ($\rho(\beta)=$const) for a homogeneous potential, so that neither RJ condensation, nor BE condensation can take place in the thermodynamic limit in 2D, i.e., the macroscopic population of the fundamental mode is not a consequence of a phase transition.
For this reason, step-index MMFs with a uniform trapping potential were not of particular interest for our study.

We have shown that, in general, when the number of photons greatly exceeds the number of available modes ($N \gg M$), the condensation properties are described by the classical RJ distribution of optical waves, while quantum condensation effects start to play a role when the number of photons decreases and approaches the number of modes. 
This is also supported by the experiments realized in MMFs that have reported the observation of RJ condensation \cite{PRL20,mangini22,pourbeyram22}.
Indeed, the number of photons involved in the experiments is related to the optical power $P$ by $ N/\tau \simeq P/(\hbar \omega_0)$, where $\omega_0$ is the central frequency of the quasi-monochromatic laser beam.
When long (typically ns) optical pulses are launched in the fiber, the time duration $\tau$ can be reasonably chosen to be of the order of the effective time duration over which the photons can interact with each other throughout their propagation in the fiber, $\tau \simeq L/(v_{g,max}-v_{g,min})$, where $v_{g,max}$ ($v_{g,min}$) are the group-velocities of the highest (lowest) mode, and $L$ the fiber length.
RJ thermalization and condensation has been demonstrated in \cite{PRL20} with $P=7$kW at the laser wavelength $\lambda=1.06 \mu$m and a MMF of $L=12$m supporting $M \simeq 120$ modes ($G \simeq 15$), which gives $\tau \simeq 2$ps, and $N \sim 10^{11}$ photons.
The experiment is thus clearly in the RJ regime, as the number of photons $N$ far outweighs the number of modes $M$.
Note that $N/M$ is exceedingly large, so that the conclusion would remain unchanged by considering different effective times scales $\tau$. 
This indicates that the investigation of photon condensation, as described by the genuine quantum BE distribution should require a different experimental set-up, with a larger amount of modes, or a smaller number of photons, and larger nonlinear interaction coefficients, so as to increase the rate of thermalization to equilibrium.

On the other hand, it has been pointed out that, for higher-order modes, the output mode power distribution that characterizes the beam self-cleaning experiments of \cite{mangini22,pourbeyram22} can be properly fitted by the BE law, see Ref.\cite{zitelli23} and Ref.\cite{zitelli24} (supplementary notes), respectively. This has been confirmed by recent experiments of mode power redistribution as a consequence of beam propagation in long spans (up to 1km) of graded-index fibers, in the presence of a significant temporal walk-off, and random coupling among the fiber modes \cite{zitelli23,zitelli24}.
These experiments realized with ultra-short pump pulses (sub-ps regime) then involve complex (2D+1) spatio-temporal effects, whose detailed description goes beyond the purely 2D spatial thermodynamic approach developed in the present work.

As a future perspective, it would also be interesting to carry out optical experiments aimed at studying the transition to condensation beyond the weakly nonlinear regime explored so far in MMFs \cite{PRL20,mangini22,pourbeyram22,podivilov22,ferraro24}.
In the strongly nonlinear regime, the condensate evolves as a distinct phase from the thermal uncondensed component, exhibiting unique physical properties such as the superfluidity of Bogoliubov sound waves, or the turbulent dynamics related to quantized vortices, as recently studied  in BE atomic condensates \cite{bagnato16}. 
For instance, considering the 2D defocusing regime of a fluid of light, it would be interesting to study the crossover between the condensation transition and the Berezinskii-Kosterlitz-Thouless transition associated to vortex-antivortex pairing.

We finally note that another important aspect has not been addressed in this paper, namely the dynamics of the out-of-equilibrium process that drives the beam of light to thermal equilibrium. 
As mentioned above, nonequilibrium thermalization is described in detail by the wave turbulence theory in the weakly nonlinear regime \cite{zakharov92,newell01,nazarenko11,Newell_Rumpf}. 
As a matter of fact, light propagation in an optical fiber is known to be affected by a structural disorder due to refractive index fluctuations introduced by inherent imperfections and environmental
perturbations \cite{mecozzi12a,mecozzi12b,mumtaz13,xiao14,psaltis16,caramazza19}. 
In this way, the observation of RJ thermalization in MMFs has stimulated the development of a wave turbulence theory that takes into account for the impact of disorder \cite{PRL19,PRA19,PRL22,kottos23,kottos24} -- also see \cite{biasi21,biasi23}.
This  revealed in particular that a `time-'dependent weak disorder can significantly increase the rate of thermalization to RJ equilibrium \cite{PRL19,PRA19}, while strong disorder can inhibit RJ thermalization \cite{PRL22}.
These works were conducted in the framework of purely classical nonlinear waves.
Consequently, the development of quantum photon fluid experiments aimed at observing BE condensation would hold fundamental significance, particularly in connection with the nonequilibrium many-body physics of disordered quantum Bose gases \cite{cherroret15,scoquart20,scoquart21,haldar23,makris24,wang20,conti17,conti22}.

\section{Appendix: ${\tilde C}_V$ for ${\tilde T} \to \infty$ in the regime $N/G^2 \ll 1$}

Bose-Einstein integrals with a frequency cut-off may be treated as Debye integrals. 
Let us define 
\begin{equation}
{\mathcal I}_s( \omega,z)= \int_0^\omega \frac{x^s}{z^{-1} e^{\frac{V_0}{\tilde T} x} -1} \, dx,
\end{equation}
then, by expanding  the denominator in a harmonic series of the fraction for $z e^{-V_0x/\tilde T} \ll 1 $:
$$ 
\frac{1}{z^{-1} e^{V_0x/\tilde T} -1} = \sum_{n=1}^\infty z^n e^{ - n\frac{V_0}{\tilde T} x}.
$$
The Debye integral becomes 
\begin{equation*}
{\mathcal I}_s( \omega,z) 
=   \sum_{n=1}^\infty z^n  \int_0^\omega {x^s} e^{ - n\frac{V_0}{\tilde T} x} \, dx,
\end{equation*}
which can easily  be integrated. 
In particular, in the limit $\omega \to \infty$:
\begin{equation*}
\lim_{\omega \to\infty} {\mathcal I}_s( \omega,z) 
=  s! ({\tilde T}/V_0)^{s+1}  g_{s+1}(z),
\end{equation*}
where $g_s(z)$
is the previously defined Bose function. 

For a finite number of modes, the energy in Eq.(\ref{eq:E_z}) and the number of particles in Eq.(\ref{eq:N_int_x_basic}) read:
\begin{align*}
{\tilde E}&=  V_0G^2 {\mathcal I}_2( 1,z)\\  
&= \frac{ \tilde TG^2}{V_0^2}   \sum_{n=1}^\infty \frac{z^n}{n^3}  \left(  2 \tilde T^2 -  \left( \tilde T^2+ (\tilde T+n V_0)^2\right)  e^{-\frac{n V_0}{\tilde T}}  \right), \\
N&= G^2 {\mathcal I}_1( 1,z)\\ 
&=  \frac{\tilde T G^2 }{V_0^2}  \sum_{n=1}^\infty \frac{ z^n}{n^2} \left(\tilde T- (\tilde T+n V_0) e^{-\frac{n V_0}{\tilde T}} \right).
\end{align*}
By expanding for $z \ll 1$ and $V_0/\tilde{T}\ll1$, we get
\begin{align*}
{\tilde E} 
&\simeq G^2 {V_0} z \Big[ \frac{1}{3}  -\frac{V_0}{4 \tilde T}  + \mathcal {O} \big( \frac{V_0^2}{\tilde T^2} \big) 
+
\mathcal {O} \big(z\big) \Big]
, 
\nonumber\\ 
N
&\simeq G^2 z \Big[ \frac{1}{2}-\frac{  {V_0} }{3 \tilde T}+ \mathcal {O} \big( \frac{V_0^2}{\tilde T^2} \big) 
+
\mathcal {O} \big(z\big) \Big]
.
\end{align*}
Note that only the first terms $n=1$ are kept in the expansions, which is equivalent to consider the classical particle limit where the Bose distribution recovers the Boltzmann distribution.
Hence, the relevant thermodynamic quantities are:
\begin{align}
&z \simeq   \frac{N}{G^2} \Big[2 + \frac{4 V_0}{3 \tilde T} + \mathcal {O} \big( \frac{V_0^2}{\tilde T^2} \big) 
+
\mathcal {O} \big(\frac{N}{G^2} \big) \Big] 
,
\label{eq:z_app}\\
&{\tilde E} \simeq   
 NV_0\Big[ \frac{2 }{3} - \frac{V_0}{18 \tilde T}  + \mathcal {O} \big( \frac{V_0^2}{\tilde T^2} \big) 
+
\mathcal {O} \big(\frac{N}{G^2} \big) \Big] 
, \\
&{\tilde C}_V \simeq  
\frac{N V_0^2}{18 \tilde T^2} 
\Big[ 1+\mathcal {O} \big( \frac{V_0}{\tilde T} \big) 
+
\mathcal {O} \big(\frac{N}{G^2} \big) \Big] .
\label{eq:Cv_app}
\end{align}
Note that the fugacity does not vanish at high temperatures, $z \to 2N/G^2$ as $\tilde T \to \infty$.
Also note that we have assumed $z \ll 1$, so that Eqs.(\ref{eq:z_app}-\ref{eq:Cv_app}) are valid in the regime $N/G^2 \ll 1$.



\begin{thebibliography}{99}
\bibitem{note_dos}
In the continuous limit we have 
$$\sum_{0\le p_x+p_y < G} \to \int_0^G dp_x \int_0^{G-p_x}dp_y=\int_0^{V_0} d\beta \rho(\beta),$$ 
where the computation of the density of states 
$$
\rho(\beta)=\int_0^G dp_x \int_0^{G-p_x}dp_y \delta\big(\beta-\beta_0(p_x+p_y)\big),
$$ 
gives Eq.(\ref{eq:dos}).
\bibitem{note_beta_ls}
The discrete sums over the modes in Eq.(\ref{eq:N_frac_bey_TL}) are carried out by considering the propagation constants associated to a circular homogeneous trapping potential, $\beta_{l,s} = x_{l,s}^2/(2 k_0 R^2)$, with $x_{l,s}$ the $s-$th zero of the Bessel function of the first kind, $J_l(x)$. For concreteness, in the plot of Fig.~\ref{fig:homog} we have considered the same value of $V_0=181 \, 346$m$^{-1}$ that is considered in Figs.~\ref{fig:0}-\ref{fig:0_BE} for the parabolic potential.
\end{thebibliography}

\end{document}